\documentclass[journal]{IEEEtran}

\IEEEoverridecommandlockouts                              
\usepackage{graphics} 
\usepackage{epsfig} 
\usepackage{mathptmx} 
\usepackage{times} 
\usepackage{amsmath} 
\usepackage{amssymb}  
\usepackage{multirow}
\usepackage{algorithm}
\usepackage{algorithmic}
\usepackage{epstopdf} 
\usepackage{stfloats} 
\usepackage{caption}
\usepackage{bm}
\usepackage{color}

\newtheorem{theorem}{\indent Theorem}

\newtheorem{definition}[theorem]{\indent Definition}

\newtheorem{remark}{\indent Remark}

\title{\LARGE \bf
Multi-agent Reinforcement Learning with Sparse Interactions by Negotiation and Knowledge Transfer}
\author{Luowei~Zhou, Pei~Yang, Chunlin~Chen,~\IEEEmembership{Member,~IEEE},
Yang~Gao,~\IEEEmembership{Member,~IEEE}
\thanks{This work was supported by the National Natural Science Foundation of China
(Nos.61273327 and 61432008).}
\thanks{L. Zhou is with the Department of Control and Systems Engineering, School of Management and Engineering, Nanjing
University, Nanjing 210093, China and with the Robotics Institute, University of Michigan, Ann Arbor, MI 48109, USA (e-mail: luozhou@umich.edu).}
\thanks{P. Yang is with the Department of Control and Systems Engineering, School of Management and Engineering, Nanjing
University, Nanjing 210093, China and with the State Key Laboratory for Novel Software Technology, Nanjing
University, Nanjing 210093, China (e-mail: yangpei@nju.edu.cn).}
\thanks{C. Chen is with the Department of Control and Systems Engineering, School of Management and Engineering, Nanjing
University, Nanjing 210093, China and with the Research Center for Novel Technology of Intelligent Equipments, Nanjing
University, Nanjing 210093, China (e-mail: clchen@nju.edu.cn).}
\thanks{Y. Gao is with the State Key Laboratory for Novel Software
Technology, Department of Computer Science, Nanjing University, Nanjing
210023, China (e-mail: gaoy@nju.edu.cn).}}

\begin{document}

\maketitle

\begin{abstract}
Reinforcement learning has significant applications for
multi-agent systems, especially in unknown dynamic environments.
However, most multi-agent reinforcement learning (MARL) algorithms
suffer from such problems as exponential computation complexity in
the joint state-action space, which makes it difficult to scale up
to realistic multi-agent problems. In this paper, a novel
algorithm named negotiation-based MARL with sparse interactions
(NegoSI) is presented. In contrast to traditional
sparse-interaction based MARL algorithms, NegoSI adopts the
equilibrium concept and makes it possible for agents to select the
non-strict Equilibrium Dominating Strategy Profile (non-strict
EDSP) or Meta equilibrium for their joint actions. The presented
NegoSI algorithm consists of four parts: the equilibrium-based
framework for sparse interactions, the negotiation for the
equilibrium set, the minimum variance method for selecting one
joint action and the knowledge transfer of local Q-values. In this
integrated algorithm, three techniques, i.e., unshared value
functions, equilibrium solutions and sparse interactions are
adopted to achieve privacy protection, better coordination and
lower computational complexity, respectively. To evaluate the
performance of the presented NegoSI algorithm, two groups of
experiments are carried out regarding three criteria: steps of
each episode (SEE), rewards of each episode (REE) and average
runtime (AR). The first group of experiments is conducted using
six grid world games and shows fast convergence and high scalability of
the presented algorithm. Then in the second group of experiments
NegoSI is applied to an intelligent warehouse problem and
simulated results demonstrate the effectiveness of the presented
NegoSI algorithm compared with other state-of-the-art MARL
algorithms.
\end{abstract}

\begin{keywords}
Knowledge transfer, multi-agent reinforcement learning, negotiation, sparse interactions.
\end{keywords}

\section{Introduction}\label{Sec1}
Multi-agent learning is drawing more and more interests from scientists and engineers in multi-agent systems (MAS) and machine learning communities \cite{Busoniu and Babuska 2008}-\cite{Hwang and Tan 2005}. One key technique for multi-agent learning is multi-agent reinforcement learning (MARL), which is an extension of reinforcement learning in multi-agent domain \cite{Yu and Zhang 2015}. Several mathematical models have been built as frameworks of MARL, such as Markov games (MG) \cite{Littman 1994} and decentralized sparse-interaction Markov decision processes (Dec-SIMDP) \cite{Melo and Veloso 2011}. Markov games are based on the assumption of full observability of all agents in the entire joint state-action space. Several well-known equilibrium-based MARL algorithms \cite{Littman 1994}-\cite{Greenwald and Hall 2003} are derived from this model. Dec-SIMDP based algorithms rely on agents' local observation, i.e., the individual state and action. Agents in Dec-SIMDP are modeled with single-agent MDP when they are outside of the interaction areas, while the multi-agent model such as MG is used when they are inside. Typical Dec-SIMDP based algorithms include LoC \cite{Melo and Veloso 2009} and CQ-learning \cite{Hauwere and Vrancx 2010}. Besides, other models such as learning automata \cite{Vrancx and Verbeeck 2008} \cite{Hauwere and Vrancx 2010-2} are also valuable tools for designing MARL algorithms.

In spite of the rapid development of MARL theories and algorithms, more efforts are needed for practical applications of MARL when compared with other MAS techniques \cite{Wang and Jiang 2014}-\cite{Stone and Veloso 2000} due to some limitations of the existing MARL methods. The equilibrium-based MARL relies on the tightly coupled learning process which hinders their applications in practice. Calculating the equilibrium (e.g., Nash equilibrium \cite{Nash 1950}) for each time step and all joint states are computationally expensive \cite{Hu and Gao 2014-1}, even for relatively small scale environments with two or three agents. In addition, sharing individual states, individual actions or even value functions all the time with other agents is unrealistic in some distributed domains (e.g., streaming processing systems \cite{An and Douglis 2008}, sensor networks \cite{An and Lesser 2011}) given the agents' privacy protections and huge real-time communication costs \cite{Busoniu and Babuska 2008}. As for MARL with sparse interactions, agents in this setting have no concept of equilibrium policy and they tend to act aggressively towards their goals, which results in a high probability of collisions.

Therefore, in this paper we focus on how the equilibrium mechanism can be used in sparse-interaction based algorithms and a negotiation-based MARL algorithm with sparse interactions (NegoSI) is proposed for multi-agent systems in unknown dynamic environments. The NegoSI algorithm consists of four parts: the equilibrium-based framework for sparse interactions, the negotiation for the equilibrium set, the minimum variance method for selecting one joint action and the knowledge transfer of local Q-values. Firstly, we start with the proposed algorithm based on the MDP model and the assumption that the agents have already obtained the single-agent optimal policy before learning in multi-agent settings. Then, the agents negotiate for pure strategy profiles as their potential set for the joint action at the ``coordination state" and they use the minimum variance method to select the relatively good one. After that, the agents move to the next joint state and receive immediate rewards. Finally, the agents' Q-values are updated by their rewards and equilibrium utilities. When initializing the Q-value for the expanded ``coordination state", the agents with NegoSI utilize both of the environmental information and the prior-trained coordinated Q-values. To test the effectiveness of the proposed algorithm, several benchmarks are adopted to demonstrate the performances in terms of the convergence, scalability, fairness and coordination ability. In addition, aiming at solving realistic MARL problems, the presented NegoSI algorithm is also further tested on an intelligent warehouse simulation platform.

The rest of this paper is organized as follows. Section \ref{Sec2} introduces the basic MARL theory and the MAS with sparse interactions. In Section \ref{Sec3}, the negotiation-based MARL algorithm with sparse interactions (NegoSI) is presented in details and related issues are discussed. Experimental results are shown in Section \ref{Sec4}. Conclusions are given in Section \ref{Sec5}.

\section{BACKGROUND}\label{Sec2}
In this section, some important concepts in multi-agent reinforcement learning and typical sparse-interaction based MARL algorithms are introduced.

\subsection{MDP and Markov Games}\label{Sec2.1}
We start with reviewing two standard decision-making models that are relevant to our work, i.e., Markov decision process (MDP) and Markov Games, respectively. MDP is the foundation of Markov Games, while Markov Games adopt game theory in multi-agent MDPs. MDP describes a sequential decision problem as follows \cite{Sutton and Barto 1998}:

\begin{definition}
(\emph{Markov Decision Process, MDP}) An MDP is a tuple $ \langle S, A, R, T \rangle$, where $S$ is the state space, $A$ is the action space of the agent, $R: S\times A\to \mathbb{R}$ is the reward function mapping state-action pairs to rewards, $T: S\times A\times S\to[0,1]$ is the transition function.
\end{definition}

An agent in an MDP is required to find an optimal policy which maximizes some reward-based optimization criteria, such as expected discounted sum of rewards:

\begin{equation}\label{Eq1}
  V^{*}(s)=\max \limits_{\pi}E_{\pi}\{\sum_{k=0}^{\infty}\gamma^{k}r^{t+k}|s^{t}=s\},
\end{equation}

\noindent where $V^{*}(s)$ stands for the value of a state $s$ under the optimal policy, $\pi:S\times A\to[0,1]$ stands for the policy of an agent, $E_{\pi}$ is the expectation under policy $\pi$, $t$ is any time step, $k$ represents a future time step, $r^{t+k}$ denotes the reward at the time step $(t+k)$ and $\gamma\in[0,1]$ is a parameter called the discount factor. This goal can also be equivalently described using the Q-value for a state-action pair:

\begin{equation}\label{Eq2}
  Q^{*}(s,a)=r(s,a)+\gamma\sum_{s'}T(s,a,s')\max \limits_{a'}Q^{*}(s',a'),
\end{equation}

\noindent where $Q^{*}(s,a)$ stands for the value of a state-action pair $(s,a)$ under the optimal policy, $s'$ is the next state and $r(s,a)$ is the immediate reward when agent adopts the action $a$ at the state $s$, $T(s,a,s')$ is the transition possibility for the agent to transit from $s$ to $s'$ given action $a$. One classic RL algorithm for estimating $Q^{*}(s,a)$ is Q-learning \cite{Watkins 1989}, whose one-step updating rule is as follows:

\begin{equation}\label{Eq3}
  Q(s,a)\leftarrow(1-\alpha)Q(s,a)+\alpha[r(s,a)+\gamma\max \limits_{a'}Q(s',a')],
\end{equation}

\noindent where $Q(s,a)$ denotes the state-action value function at a state-action pair $(s,a)$ and $\alpha\in[0,1]$ is a parameter called the learning rate. Provided that all state-action pairs are visited infinite times with a reasonable learning rate, the estimated Q-value $Q(s,a)$ converges to $Q^{*}(s,a)$ \cite{Tsitsiklis 1994}.

Markov games are widely adopted as a framework for multi-agent reinforcement learning (MARL) \cite{Littman 1994} \cite{Hu and Wellman 2003}. It is regarded as multiple MDPs in which the transition probabilities and rewards depend on the joint state-action pairs of all agents. In a certain state, agents' individual action sets generate a repeated game that could be solved in a game-theoretic way. Therefore, Markov game is a richer framework which generalizes both of the MDP and the repeated game \cite{Nowe and Vrancx 2012}-\cite{Burkov 2010}.

\begin{definition}
(\emph{Markov game}) An $n$-agent ($n\ge2$) Markov game is a tuple $\langle n, \{S_{i}\}_{i=1,\ldots,n}, \{A_{i}\}_{i=1,\ldots,n}, \{R_{i}\}_{i=1,\ldots,n}, T \rangle$, where $n$ is the number of agents in the system, $S_i$ is the set of the state space for $i^{th}$ agent, $S=\{S_{i}\}_{i=1,\ldots,n}$ is the set of state spaces for all agents, $A_i$ is the set of the action space for $i^{th}$ agent, $A=\{A_{i}\}_{i=1,\ldots,n}$ is the set of action spaces for all agents, $R_{i}:S\times A\to\mathbb{R}$ is the reward function of agent $i$, $T:S\times A\times S\to[0,1]$ is the transition function.
\end{definition}

Denote the individual policy of agent $i$ by $\pi_{i}=S\times
A_{i}\to[0,1]$ and the joint policy of all agents by
$\pi=(\pi_{1},\ldots,\pi_{n})$. The Q-value of the join
state-action pair for agent $i$ under the joint policy $\pi$ can
be formulated by

\begin{equation}\label{Eq4}
  Q_{i}^{\pi}(\vec{s},\vec{a})=E_{\pi}\{\sum_{k=0}^{\infty}\gamma^{k}r_{i}^{t+k}|\vec{s}^{t}=\vec{s},\vec{a}^{t}=\vec{a}\},
\end{equation}

\noindent where $\vec{s}\in S$ stands for a joint state, $\vec{a}\in A$ for a joint action and $r_{i}^{t+k}$ is the reward received at the time step $(t+k)$. Unlike the optimization goal in an MDP, the objective of a Markov game is to find an equilibrium joint policy $\pi$ rather than an optimal joint policy for all agents. Here, the equilibrium policy concept is usually transferred to finding the equilibrium solution for the one-shot game played in each joint state of a Markov game \cite{Hu and Gao 2014-1}. Several equilibrium-based MARL algorithms in existing literatures such as NashQ \cite{Hu and Wellman 2003} \cite{Porter and Nudelman 2008} and NegoQ \cite{Hu and Gao 2014-1} have been proposed, so that the joint state-action pair Q-value can be updated according to the equilibrium:

\begin{equation}\label{Eq5}
  Q_{i}(\vec{s},\vec{a})\leftarrow(1-\alpha)Q_{i}(\vec{s},\vec{a})+\alpha(r_{i}(\vec{s},\vec{a})+\gamma\phi_{i}(\vec{s}'))),
\end{equation}

\noindent where $\phi_{i}(\vec{s}')$ denotes the expected value of the equilibrium in the next joint state $\vec{s}'$ for agent $i$ and can be calculated in the one-shot game at that joint state.

\subsection{MAS with Sparse Interactions}\label{Sec2.2}
The definition of Markov game reveals that all agents need to learn their policies in the full joint state-action space and they are coupled with each other all the time \cite{Hu and Wellman 2003} \cite{Hu and Gao 2015}. However, this assumption does not hold in practice. The truth is that the learning agents in many practical multi-agent systems are loosely coupled with some limited interactions in some particular areas \cite{Yu and Zhang 2015} \cite{Melo and Veloso 2009} \cite{Hu and Gao 2015}. Meanwhile, the interactions between agents may not always involve all the agents. These facts lead to a new mechanism of sparse interactions for MARL research.

\begin{figure}
  \centering
  \includegraphics[width=2.2in]{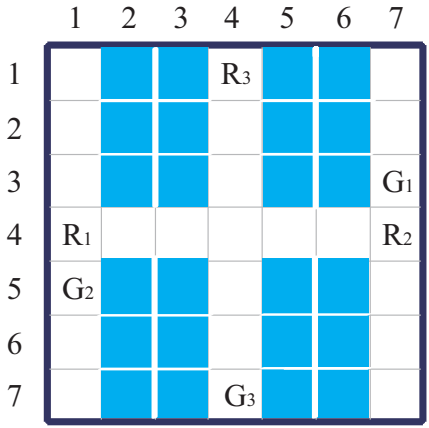}
  \caption{A part of an intelligent warehouse with three robots, where $R_i$ ($i=1,2,3$) represents for robot $i$ with its corresponding goal $G_i$ and the shaded grids are storage shelves.}\label{fig1}
\end{figure}

Sparse-interaction based algorithms \cite{Melo and Veloso 2009} \cite{Hauwere and Vrancx 2010} have recently found wide applications in MAS research. An example of MAS with sparse interactions is the intelligent warehouse systems, where autonomous robots only consider other robots when they are close enough to each other \cite{Nowe and Vrancx 2012} (see Fig. \ref{fig1}), i.e., when they meet around the crossroad. Otherwise, they can act independently.

Earlier works such as the coordinated reinforcement learning \cite{Guestrin and Lagoudakis 2002} \cite{Guestrin and Venkataraman 2002} and sparse cooperative Q-learning \cite{Kok and Vlassis 2004} used coordination graphs (CGs) to learn interdependencies between agents and decomposed the joint value function to local value functions. However, these algorithms cannot learn CGs online and only focus on finding specific states where coordination is necessary instead of learning for coordination \cite{Melo and Veloso 2009}. Melo and Veloso \cite{Melo and Veloso 2009} extended the Q-learning to a two-layer algorithm and made it possible for agents to use an additional COORDINATE action to determine the state when the coordination was necessary. Hauwere and Vrancx proposed Coordinating Q-learning (CQ-learning) \cite{Hauwere and Vrancx 2010} and FCQ-learning \cite{Hauwere and Vrancx 2012} that helped agents learn from statistical information of rewards and Q-values where an agent should take other agents' states into account. However, all these algorithms allow agents to play greedy strategies at a certain joint state rather than equilibrium strategies, which might cause conflicts.

More recently, Hu et al \cite{Hu and Gao 2014-1} proposed an efficient equilibrium-based MARL method, called Negotiation-based Q-learning, by which agents can learn in a Markov game with unshared value functions and unshared joint state-actions. In later work, they applied this method for sparse interactions by knowledge transfer and game abstraction \cite{Hu and Gao 2015}, and demonstrated the effectiveness of the equilibrium-based MARL in solving sparse-interaction problems. Nevertheless, as opposed to single Q-learning based approaches like CQ-learning, Hu's equilibrium-based methods for sparse interactions require a great deal of real-time information about the joint states and joint actions of all the agents, which results in huge amount of communication costs. In this paper, we focus on solving learning problems in complex systems. Tightly coupled equilibrium-based MARL methods discussed above are impractical in these situations while sparse-interaction based algorithms tend to cause many collisions. To this end, we adopt the sparse-interaction based learning framework and each agent selects equilibrium joint actions when they are in coordination areas.

\section{NEGOTIATION-BASED MARL WITH SPARSE INTERACTIONS}\label{Sec3}
When people work in restricted environments with possible conflicts, they usually learn how to finish their individual tasks first and then learn how to coordinate with others. We apply this commonsense to our sparse-interaction method and decompose the learning process into two distinct sub-processes \cite{Yu and Zhang 2015}. First, each agent learns an optimal single-agent policy by itself in the static environment and ignores the existences of other agents. Second, each agent learns when to coordinate with others according to their immediate reward changes, and then learns how to coordinate with others in a game-theoretic way. In this section, the negotiation-based framework for MAS with sparse interactions is first introduced and then related techniques and specific algorithms are described in details.

\subsection{Negotiation-based Framework for Sparse Interactions}\label{Sec3.1}
We assume that agents have learnt their optimal policies and
reward models when acting individually in the environment. Two
situations might occur when agents are working in a multi-agent
setting. If the received immediate rewards for state-action pairs
are the same as what they learned by reward models, the agents act
independently. Otherwise, they need to expand
their individual state-action pairs to the joint ones by adding
other agents' state-action pairs for better coordination. This
negotiation-based framework for sparse interactions is given as
shown in Algorithm \ref{tab:algorithm1}, while the expansion and
negotiation process is explained as follows:

\begin{algorithm*}[!hb]
\begin{algorithmic}[1]
\caption{Negotiation-based Framework for Sparse Interactions}\label{tab:algorithm1}
\REQUIRE The agent $i$, state space $S_i$, action space $A_i$, learning rate $\alpha$, discount rate $\gamma$, exploration factor $\epsilon$ for the $\epsilon-Greedy$ exploration policy.\\
\ENSURE Global Q-value $Q_i$ with optimal single-agent policy.\\
\FOR{\textbf{each} episode}
    \STATE Initialize state $s_i$;
    \WHILE{true}
        \STATE Select $a_i\in A_i$ from $Q_i$ with $\epsilon-Greedy$;
        \IF {($s_i,a_i$) is ``dangerous"}
            \STATE Broadcasts ($s_i,a_i$) and receives ($s_{-i},a_{-i}$), form ($\vec{s},\vec{a}$);\\
            /* See Section \ref{Sec3.4} for the definitions of $s_{-i}$ and $a_{-i}$*/\\
            \IF {($\vec{s},\vec{a}$) is not ``coordination pair"}
                \STATE Mark ($\vec{s},\vec{a}$) as a ``coordination pair" and $\vec{s}$ as a ``coordination state", initialize Local Q-value $Q_i^J$ at $\vec{s}$ with transfer knowledge (See Equation \ref{Eq9});
            \ENDIF
            \STATE Negotiate for the equilibrium joint action with Algorithm \ref{tab:algorithm2} - Algorithm \ref{tab:algorithm4}. Select new $\vec{a}$ with $\epsilon-Greedy$;
        \ELSE [detected an immediate reward change]
            \STATE Mark ($s_i,a_i$) as ``dangerous", broadcasts ($s_i,a_i$) and receives ($s_{-i},a_{-i}$), form ($\vec{s},\vec{a}$);
            \STATE Mark ($\vec{s},\vec{a}$) as ``coordination pair" and $\vec{s}$ as ``coordination state", initialize $Q_i^J$ at $\vec{s}$ with transfer knowledge (See Equation \ref{Eq9});
            \STATE Negotiate for the equilibrium joint action with Algorithm \ref{tab:algorithm2} - Algorithm \ref{tab:algorithm4}. Select new $\vec{a}$ with $\epsilon-Greedy$;
        \ENDIF
        \STATE Move to the next state $s_i'$ and receive the reward $r_i$;
        \IF {$\vec{s}$ exists}
            \STATE Update $Q_i^J(\vec{s},\vec{a})=(1-\alpha)Q_i^J(\vec{s},\vec{a})+\alpha(r_i+\gamma\max\limits_{a_i'}Q_i(s_i',a_i'))$;
        \ENDIF
        \STATE Update $Q_i(s_i,a_i)=(1-\alpha)Q_i(s_i,a_i)+\alpha(r_i+\gamma\max\limits_{a_i'}Q_i(s_i',a_i'))$;
        \STATE $s_i\leftarrow s_i'$;
    \ENDWHILE{ \textbf{until} $s_i$ is a terminal state.}
\ENDFOR
\end{algorithmic}
\end{algorithm*}

\begin{enumerate}
\item Broadcast joint state. Agents select an action at a certain state, and they detect a change in the immediate rewards. This state-action pair is marked as ``dangerous" and these agents are called ``coordinating agents". Then, ``coordinating agents" broadcast their state-action information to all other agents and receive corresponding state-action information from others. These state-action pairs with reward changes form a joint state-action and is marked as a ``coordination pair". Also, these states form a joint state called a  ``coordination state", which is included in the state space of each ``coordinating agent".
\item Negotiation for equilibrium policy. When agents select a ``dangerous" state-action pair, they broadcast their state-action information to each other to determine whether they are staying at a ``coordination pair". If so, the agents need to find an equilibrium policy rather than their inherent greedy policies to avoid collisions. We propose a negotiation mechanism similar to the work in \cite{Hu and Gao 2014-1} to find this equilibrium policy. Each ``coordinating agent" broadcasts the set of strategy profiles that are potential Non-strict Equilibrium-Dominating Strategy Profile (non-strict EDSP) according to its own utilities (See Algorithm \ref{tab:algorithm2}). If no non-strict EDSP is found, the agents search for a Meta equilibrium set (See Algorithm \ref{tab:algorithm3}) instead, which is always nonempty \cite{Hu and Gao 2014-1}. Then, if there are several candidates in the equilibrium set, the agents use the minimum variance method to find the relatively good one (See Algorithm \ref{tab:algorithm4}).
\item When an agent selects an action at a certain state, if neither a change in the immediate reward nor a ``dangerous" state-action pair is detected, the agent continue to select its action independently.
\end{enumerate}

\begin{figure*}
  \centering
  \includegraphics[width=6in]{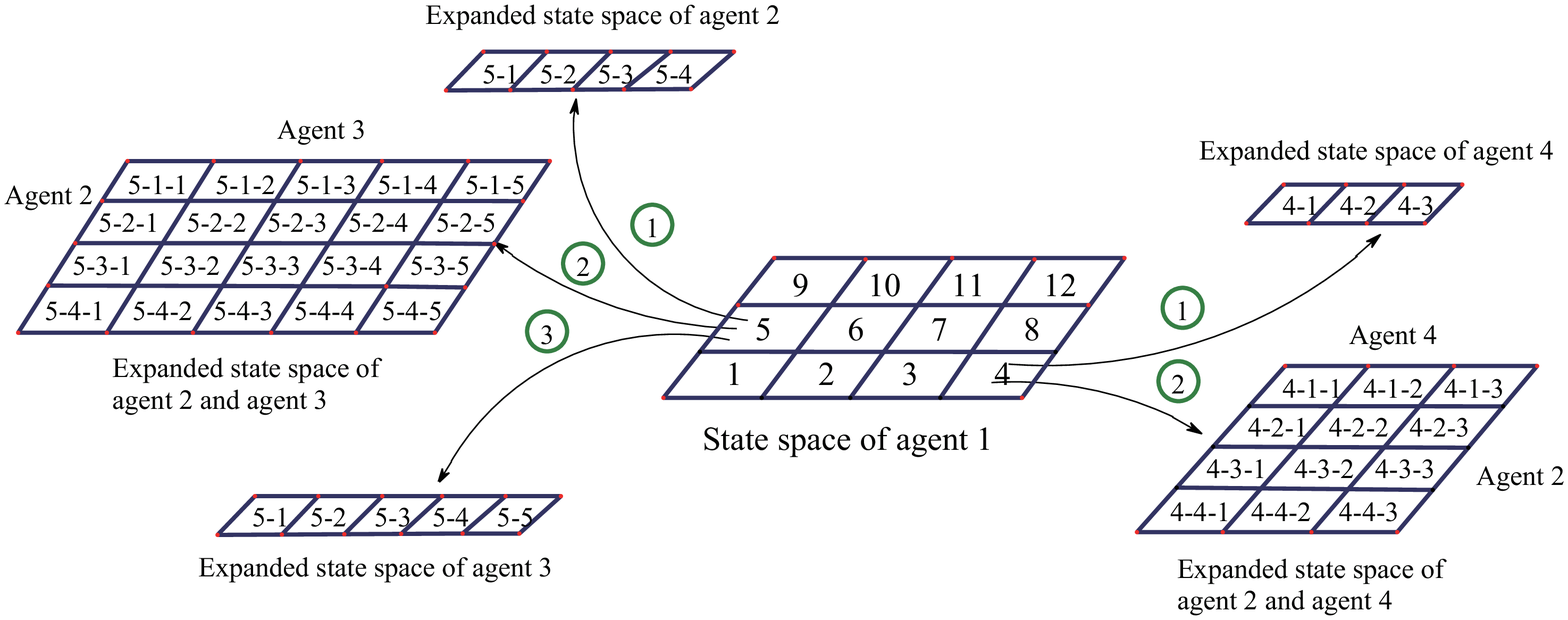}
  \caption{An example of the expansion of an agent's state space.}\label{fig1plus}
\end{figure*}

\begin{remark}
If the number of agents $n\ge2$, the agents may have several
different expansions for different ``coordination states" in one
state. Each expanded joint-state is assigned an index and has a
corresponding local Q-value. Once the agents form a ``coordination
state", they search for the joint state in their expanded
joint-state pool and obtain the corresponding local Q-value. An
example is shown as in Fig. \ref{fig1plus}. In this example, agent
1-4 have twelve states, four states, five states and three states,
respectively. For convenience, we assume here when
agent 1 is expanding joint states, it adds all states of relevant
agents in its joint states even though in most of the cases it
only needs to add a small part of the states. Specifically, the
fourth state of agent 1 is expanded to two joint states, the first
one is with agent 4 and the second one is with agent 2 and agent
4. Similarly, the fifth state of agent 1 is expanded to three
joint states, the first one is with agent 2, the second one is
with agent 2 and agent 3, and the third one is with agent 3.
\end{remark}

\begin{remark}
Owing to the broadcast mechanism, an agent will add the states of
all the ``coordinating agents" to its state space, even though it
does not need to coordinate with some of these agents. The agents
cannot distinguish the agent they need to coordinate with from
other agents, which brings unnecessary calculations. This problem
becomes more significant with the growth of the complexity of the
environment and the number of the agents. Our tentative solution
is setting the communication range of the agents so that they can
only add neighbouring agents' states to their state space and
ignore those that do not need to coordinate with.
\end{remark}

\begin{remark}
It is worth noting that if the agent detects a
``dangerous'' state-action pair, it should observe a reward
change, that is, if we replace the former concept with the latter
one, the algorithm can work equally. Defining dangerous
state-action pairs mainly help us to better explain our thoughts
and describe the algorithm.
\end{remark}

\vspace{-10pt}

\subsection{Negotiation for Equilibrium Set}\label{Sec3.2}
One contribution of this paper is to apply the equilibrium
solution concept to traditional Q-learning based MARL algorithms.
Different from previous work like CQ-learning and Learning of
Coordination \cite{Hauwere and Vrancx 2010}, our approach aims at
finding the equilibrium solution for the one-shot game played in
each ``coordination state". As a result, we focus on two pure
strategy profiles \cite{Hu and Gao 2014-1}, i.e., Non-strict
Equilibrium-Dominating Strategy Profile (non-strict EDSP) and Meta
Equilibrium set. The definition of Non-strict EDSP is as follows.

\begin{definition}
(\emph{Non-strict EDSP}): In an $n$-agent ($n\ge2$) normal-form game $\Gamma$, $\vec{e_{i}}\in A (i=1,2,\ldots,m)$ are pure strategy Nash equilibriums. A joint action $\vec{a}\in A$ is a non-strict EDSP if $\forall j\le n$,
\begin{equation}\label{Eq6}
  U_{j}(\vec{a})\ge\min\limits_{i}U_{j}(\vec{e_{i}})
\end{equation}
\end{definition}

The negotiation process of finding the set of Non-strict EDSP is shown as in Algorithm \ref{tab:algorithm2}, which generally consists of two steps: 1) the agents find the set of strategy profiles that are potentially Non-strict EDSP according to their individual utilities; 2) the agents solve the intersection of all the potentially strategy sets and get the Non-strict EDSP set.

However, given the fact that the pure strategy Nash equilibrium
can be non-existent in some cases, the set of Non-strict EDSP
might also be empty. On this occasion, we replace this strategy
profile with Meta Equilibrium, which is always nonempty. The
sufficient and necessary condition of Meta Equilibrium is defined
as follows.

\begin{algorithm}[!bp]
\caption{Negotiation for Non-strict EDSPs Set}\label{tab:algorithm2}
\begin{algorithmic}[1]
\REQUIRE A normal-form game $ \langle n,\{A_i\}_{i=1,\ldots,n},\{U_i\}_{i=1,\ldots,n} \rangle $.\\
/* To be noted, ``coordinating agent" $i$ only has the knowledge of $n$, $\{A_i\}_{i=1,\ldots,n}$ and $U_i$*/\\
\ENSURE The set of non-strict EDSP candidates for ``coordinating agents" $i$: $J_{NS}^i\leftarrow\emptyset$;\\
Minimum utility value of pure strategy Nash equilibrium (PNE) candidates for ``coordinating agents" $i$: $MinU_{PNE}^i\leftarrow\infty$;\\
The set of non-strict EDSPs: $J_{NS}\leftarrow\emptyset$.\\
\FOR{\textbf{each} $\vec{a_{-i}}\in A_{-i}$}
    \IF {$\max\limits_{a_i'}U_i(a_i',\vec{a_{-i}})<MinU_{PNE}^i$}
        \STATE $MinU_{PNE}^i=\max\limits_{a_i'}U_i(a_i',\vec{a_{-i}})$;
    \ENDIF
\ENDFOR
\FOR{\textbf{each} $\vec{a}\in A$}
    \IF {$U_i(\vec{a})\ge MinU_{PNE}^i$}
        \STATE $J_{NS}^i\leftarrow J_{NS}^i \cup \{\vec{a}\}$;
    \ENDIF
\ENDFOR\\
/* Broadcast $J_{NS}^i$ and corresponding utilities*/\\
\STATE $J_{NS}\leftarrow \bigcap_{i=1}^{n}J_{NS}^i$
\end{algorithmic}
\end{algorithm}

\begin{definition}
(\emph{Sufficient and Necessary Condition of Meta Equilibrium})
\cite{Hu and Gao 2014-1}: In an $n$-agent ($n\ge2$) normal-form
game $\Gamma$, a joint action $\vec{a}$ is called a Meta
equilibrium from a metagame $k_{1}k_{2}\ldots k_{r}\Gamma$ if and
only if for any $i$ there holds
\begin{equation}\label{Eq7}
  U_i(\vec{a})\ge \min\limits_{\vec{a_{P_i}}}\max\limits_{a_i}\min\limits_{\vec{a_{S_i}}}U_i(\vec{a_{P_i}},a_{i},\vec{a_{S_i}}),
\end{equation}
\end{definition}

\noindent where $P_i$ is the set of agents listed before sign $i$ in the prefix $k_1k_2\ldots k_r$, $S_i$ is the set of agents listed after sign $i$ in the prefix.

For example, in a three agents metagame $213\Gamma$, we have $P_1=\{2\}, S_1=\{3\}; P_2=\emptyset, S_2=\{1,3\}; P_3=\{2,1\},S_3=\emptyset$. A Meta equilibrium $\vec{a}$ meets the following constraints:

\begin{equation}\label{Eq8}
\begin{split}
  U_1(\vec{a})\ge \min\limits_{a_2}\max\limits_{a_1}\min\limits_{a_3}U_1(a_1,a_2,a_3),\\
  U_2(\vec{a})\ge \max\limits_{a_2}\min\limits_{a_1}\min\limits_{a_3}U_2(a_1,a_2,a_3),\\
  U_3(\vec{a})\ge \min\limits_{a_2}\min\limits_{a_1}\max\limits_{a_3}U_3(a_1,a_2,a_3).
\end{split}
\end{equation}

Hu et al \cite{Hu and Gao 2014-1} used a negotiation-based method to find Meta Equilibrium set and we simplified the method as shown in Algorithm \ref{tab:algorithm3}. It is also pointed out that both of these two strategy profiles belong to the set of symmetric meta equilibrium, which to some degree strengthens the convergence of the algorithm.

\begin{algorithm}[!hb]
\caption{Negotiation for Meta Equilibrium Set for 3-agent games}\label{tab:algorithm3}
\begin{algorithmic}[1]
\REQUIRE A normal-form game $<n,\{A_i\}_{i=1,\ldots,n},\{U_i\}_{i=1,\ldots,n}>$.\\
/* To be noted, ``coordinating agents" $i$ only has the knowledge of $n$, $\{A_i\}_{i=1,\ldots,n}$ and $U_i$*/\\
\ENSURE The set of Meta Equilibrium candidates for ``coordinating agents" $i$: $J_{MetaE}^i\leftarrow\emptyset$;\\
Minimum utility value of Meta Equilibrium candidates for ``coordinating agents" $i$: $MinU_{MetaE}^i\leftarrow\infty$;\\
The set of Meta Equilibrium: $J_{MetaE}\leftarrow\emptyset$;\\
\STATE Randomly initialize the prefix $s_1s_2s_3$ from the set \{123,132,213,231,312,321\}.\\
\STATE Calculate $MinU_{MetaE}^i$ according to Equation \ref{Eq7};
\FOR{\textbf{each} $\vec{a}\in A$}
    \IF {$U_i(\vec{a})\ge MinU_{MetaE}^i$}
        \STATE $J_{MetaE}^i\leftarrow J_{MetaE}^i \cup \{\vec{a}\}$;
    \ENDIF
\ENDFOR\\
/* Broadcast $J_{MetaE}^i$ and corresponding utilities*/\\
\STATE $J_{MetaE}\leftarrow \bigcap_{i=1}^{n}J_{MetaE}^i$
\end{algorithmic}
\end{algorithm}

\subsection{Minimum Variance Method}\label{Sec3.3}
In Section \ref{Sec3.2}, we presented the process for all ``coordinating agents" to negotiate for an equilibrium joint-action set. However, the obtained set usually contains many strategy profiles and it is difficult for the agents to choose an appropriate one. In this paper a Minimum Variance method is proposed to help the ``coordinating agents" choose the joint action with relatively high total utility and minimum utilities variance to guarantee the cooperation and fairness of the learning process. In addition, if the equilibrium set is nonempty, the best solution defined in the minimum variance method always exists. The minimum variance method is described in Algorithm \ref{tab:algorithm4}.

\begin{algorithm}[!ht]
\caption{Minimum Variance Method}\label{tab:algorithm4}
\begin{algorithmic}[1]
\REQUIRE The equilibrium set with $m$ elements $J_{NS}=\{\vec{a_{ns}^1},\vec{a_{ns}^2},\ldots,\vec{a_{ns}^m}\}$ and corresponding utilities $\{U_i^{ns}\}_{i=1,\ldots,n}$.
\ENSURE threshold value for total utility $\tau$;\\
/* We set the threshold value to the mean value of the sum utilities of different joint-action profiles $\frac{\sum_{j=1}^m\sum_{i=1}^nU_i^{ns}(\vec{a_{ns}^j})}{m}$*/\\
Minimum variance of these equilibriums $MinV\leftarrow\infty$;\\
Best equilibrium of the non-strict EDSPs set $J_{BestNS}\leftarrow\emptyset$.\\
\FOR{\textbf{each} $\vec{a_{ns}^j}\in J_{NS}$}
    \IF {$\sum_{i=1}^{n}U_{i}^{ns}(\vec{a_{ns}^j})<\tau$}
        \STATE $J_{NS}\leftarrow J_{NS}\backslash\{\vec{a_{ns}^j}\}$;
    \ENDIF
\ENDFOR\\
/* Minimize the joint action's utility variance*/
\FOR{\textbf{each} $\vec{a_{ns}^j}\in J_{NS}$}
    \IF {$\sqrt{\frac{1}{n}\sum_{k=1}^n[U_k^{ns}(\vec{a_{ns}^j})-\frac{1}{n}\sum_{i=1}^n U_i^{ns}(\vec{a_{ns}^j})]^2}< MinV$}
        \STATE $MinV=\sqrt{\frac{1}{n}\sum_{k=1}^n[U_k^{ns}(\vec{a_{ns}^j})-\frac{1}{n}\sum_{i=1}^n U_i^{ns}(\vec{a_{ns}^j})]^2}$;\\
        \STATE $J_{BestNS}=\{\vec{a_{ns}^j}\}$;\\
    \ENDIF
\ENDFOR
\STATE \textbf{Output} $J_{BestNS}$ as the adopted joint action.\\
\end{algorithmic}
\end{algorithm}

After negotiating for one equilibrium, the agents update their states according to the joint action and receive immediate rewards, which are used to update local Q-values as well as global Q-values. Unlike other sparse-interaction algorithms (e.g., CQ-learning), we update the global optimal Q-values to avoid possible misleading information. In fact, in some cases, the selected policies in multi-agent setting are totally opposite to the agents' individual optimal policies due to dynamic coordinating processes. The whole negotiation-based learning algorithm has already been given in Algorithm \ref{tab:algorithm1}.

\subsection{Local Q-value transfer}\label{Sec3.4}
\begin{figure}[b]
  \centering
  \includegraphics[width=3in]{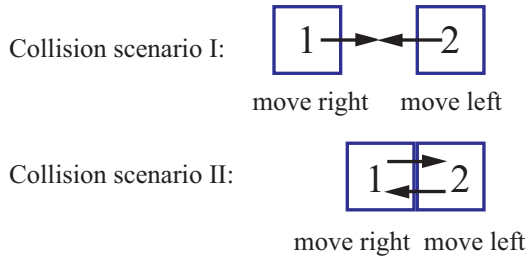}
  \caption{Two scenarios of possible collisions.}\label{fig2}
\end{figure}

At the beginning of the learning process, we use the transfer knowledge of agents' optimal single agent policy to accelerate the learning process. Furthermore, it is possible to improve the algorithm performance by the local Q-value transfer. In most previous literatures \cite{Melo and Veloso 2009} \cite{Hauwere and Vrancx 2010}, the initial local Q-values of the newly expanded joint states are zeros. Recently, Vrancx et al proposed a transfer learning method \cite{Vrancx and Hauwere 2011} to initialize these Q-values with prior trained Q-value from the source task, which is reasonable in the real world. When people meet with others on the way to their individual destinations, they usually have prior knowledge of how to avoid collisions. Based on this prior commonsense and the knowledge of how to finish their individual tasks, the agents learn to negotiate with others and obtain fixed coordination strategies suitable for certain environments. However, Vrancx et al only used coordination knowledge to initialize the local Q-values and overlooked the environmental information, which was proved to be less effective in our experiments. In our approach, we initialize the local Q-values of newly detected ``coordination" to hybrid knowledge as follows:

\begin{equation}\label{Eq9}
  Q_{i}^{J}((s_i,\vec{s_{-i}}),(a_i,\vec{a_{-i}}))\leftarrow Q_i(s_i,a_i)+Q_i^{CT}(\vec{s},\vec{a}),
\end{equation}

\noindent where $\vec{s_{-i}}$ is the joint-state set except for the state $s_i$ and $\vec{a_{-i}}$ is the joint-action set except for the action $a_i$, $Q_{i}^{J}((s_i,\vec{s_{-i}}),(a_i,\vec{a_{-i}}))$ is equal with $Q_i^{J}(\vec{s},\vec{a})$, $Q_i^{CT}(\vec{s},\vec{a})$ is the transferred Q-value from a blank source task at joint state $\vec{s}$. In this blank source task, joint action learners (JAL) \cite{Claus and Boutilier 1998} learn to coordinate with others disregarding the environmental information. They are given a fixed number of learning steps to learn stable Q-value $Q_i^{CT}(\vec{s},\vec{a})$. Similar to \cite{Vrancx and Hauwere 2011}, the joint state $\vec{s}$ in $Q_i^{CT}(\vec{s},\vec{a})$ is presented as the relative position $(\Delta x,\Delta y)$, horizontally and vertically. When agents attempt to move into the same grid location or their previous locations (as shown in Fig. \ref{fig2}), these agents receive a penalty of -10. In other cases, the reward is zero.

\begin{figure}[!tp]
\begin{minipage}{1\linewidth}\label{fig3}
  \centerline{\includegraphics[width=1.5in]{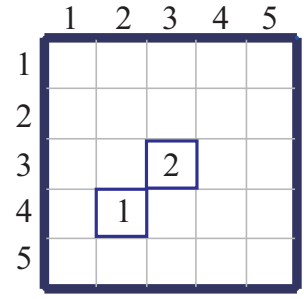}}
  \centerline{(a) A $5\times5$ blank source task with two agents.}
\end{minipage}

\vspace{8pt}
\begin{minipage}{1\linewidth}\label{fig4}
  \centerline{\includegraphics[width=2.5in]{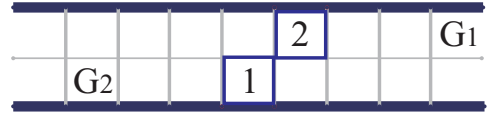}}
  \centerline{(b) The detected ``coordination states".}
\end{minipage}
\caption{An example of the local Q-value transfer in a two-agent system.}
\label{fig3plus4}
\end{figure}

Take a two-agent source task for example (as shown in Fig. \ref{fig3plus4}(a)). The locations of agent $1$ and agent $2$ are ($4, 3$) and ($3, 4$), respectively. Then we have the joint state $\vec{s}=(x_1-x_2, y_1-y_2)=(1, -1)$ and the joint action for this joint state
\begin{gather*}
\vec{a}=\begin{pmatrix}(a_1, a_2)\end{pmatrix}=
\begin{pmatrix} (\uparrow, \uparrow) & (\uparrow, \downarrow) & (\uparrow, \leftarrow) & (\uparrow, \rightarrow) \\
                (\downarrow, \uparrow) & (\downarrow, \downarrow) & (\downarrow, \leftarrow) & (\downarrow, \rightarrow) \\
                (\leftarrow, \uparrow) & (\leftarrow, \downarrow) & (\leftarrow, \leftarrow) & (\leftarrow, \rightarrow) \\
                (\rightarrow, \uparrow) & (\rightarrow, \downarrow) & (\rightarrow, \leftarrow) & (\rightarrow, \rightarrow)
\end{pmatrix},
\end{gather*}
where $\{\uparrow, \downarrow, \leftarrow, \rightarrow\}$ denotes the action set of \{\emph{up, down, left, right}\} for each agent. After sufficient learning steps, the agents learn their Q-value matrices at joint state $\vec{s}$ as

\vspace{-15pt}

\begin{gather*}
Q_1^{CT}(\vec{s},\bm\cdot)=Q_2^{CT}(\vec{s},\bm\cdot)=
\begin{pmatrix} 0 & 0 & -10 & 0 \\
                0 & 0 & 0 & 0 \\
                0 & 0 & 0 & 0 \\
                0 & -10 & 0 & 0
\end{pmatrix},
\end{gather*}

\noindent that is because when agent 1 and agent 2 select the action pair $(\uparrow, \leftarrow)$ or $(\rightarrow, \downarrow)$, a collision will occur, which leads to a -10 punishment for each agent. Otherwise, the reward is 0. Suppose that the agents are in the ``coordination state" as shown
in Fig. \ref{fig3plus4}(b). When acting independently in the
environment, the agents learn their single agent optimal Q-value
vectors at state $s_1$ or $s_2$ as
$Q_1(s_1,\bm\cdot)=(-1,-10,-5,-1)$,
$Q_2(s_2,\bm\cdot)=(-10,-1,-1,-5)$. Then the local Q-value matrices
of this ``coordination state" need to be initialized as
\begin{gather*}
\begin{aligned}
Q_1^{J}(\vec{s},\bm\cdot)&=Q_1(s_1,\bm\cdot)^T\times(1,1,1,1)+Q_1^{CT}(\vec{s},\bm\cdot)\\&=
\begin{pmatrix} -1 & -1 & -11 & -1 \\
                -10 & -10 & -10 & -10 \\
                -5 & -5 & -5 & -5 \\
                -1 & -11 & -1 & -1
\end{pmatrix},
\\
Q_2^{J}(\vec{s},\bm\cdot)&=(1,1,1,1)^T\times Q_2(s_2,\bm\cdot)+Q_2^{CT}(\vec{s},\bm\cdot)\\&=
\begin{pmatrix} -10 & -1 & -11 & -5 \\
                -10 & -1 & -1 & -5 \\
                -10 & -1 & -1 & -5 \\
                -10 & -11 & -1 & -5
\end{pmatrix}.\\
\end{aligned}
\end{gather*}

So the pure strategy Nash equilibria for this joint state are (\emph{up, down}) and (\emph{right, left}). If we initialize the local Q-value with the way used in \cite{Vrancx and Hauwere 2011} or just initialize them to zeros, the learning process would be much longer.

For the ``coordination state" with three agents, the Q-value $Q_i^{CT}(\vec{s},\vec{a})$ can be calculated in the same way, except for the relative positions $(\Delta x_1,\Delta y_1,\Delta x_2,\Delta y_2)=(x_1-x_2,y_1-y_2,x_2-x_3,y_2-y_3)$ and the cubic Q-value for each joint state.

\section{EXPERIMENTS}\label{Sec4}
To test the presented NegoSI algorithm, several groups of simulated experiments are implemented and the results are compared with those of three other state-of-the-art MARL algorithms, namely, CQ-learning \cite{Hauwere and Vrancx 2010}, NegoQ with value function transfer (NegoQ-VFT) \cite{Hu and Gao 2015} and independent learners with value function transfer (IL-VFT). In next two subsections, the presented NegoSI algorithm is applied to six grid world games and an intelligent warehouse problem, which shows that NegoSI is an effective approach for MARL problems compared with other existing MARL algorithms.

For all these experiments, each agent has four actions: up, down, left, right. The reward settings are as follows:
\begin{enumerate}
  \item When an agent reaches its goal/goals, it receives a reward of 100. Its final goal is an absorbing state. One episode is over when all agents reach their goals.
  \item A negative reward of -10 is received if a collision happens or an agent steps out of the border. In these cases, agents will bounce to their previous states.
  \item Otherwise, the reward is set to -1 as the power consumption.
\end{enumerate}
The settings of other parameters are the same for all algorithms:
learning rate $\alpha=0.1$, discount rate $\gamma=0.9$,
exploration factor $\epsilon=0.01$ for $\epsilon$-greedy strategy.
All algorithms run $2000$ iterations for the grid world games and
$8000$ iterations for the intelligent warehouse problem. We use
three typical criteria to evaluate these MARL algorithms, i.e.,
steps of each episode (SEE), rewards of each episode (REE) and
average runtime (AR). All the results are averaged over 50 runs.

\subsection{Tests on grid world games}\label{Sec4.1}

\begin{figure}[!bp]
\begin{minipage}{0.49\linewidth}\label{fig5a}
  \centerline{\includegraphics[width=1in]{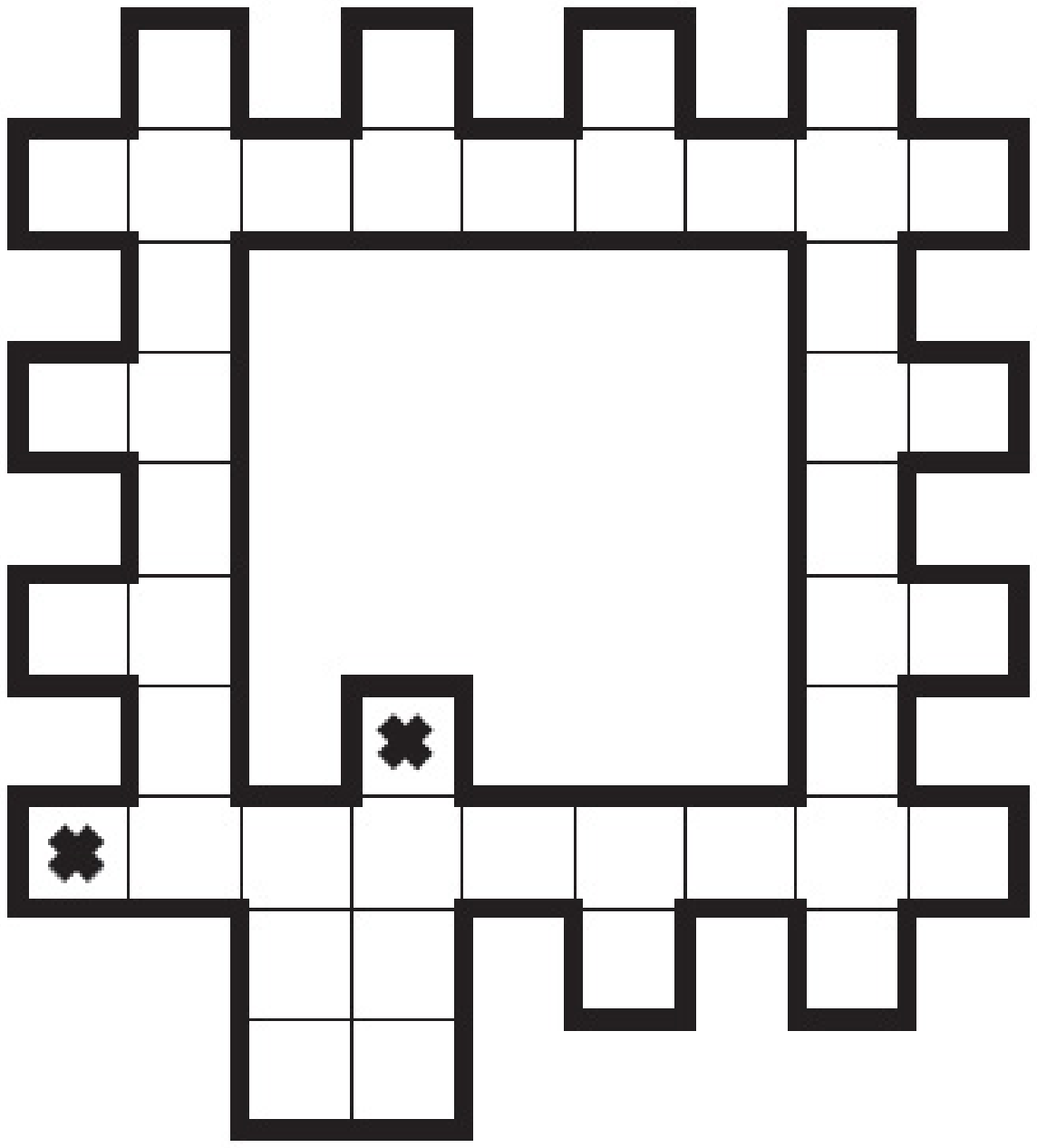}}
  \vspace{-3pt}
  \centerline{(a) ISR}
\end{minipage}
\hfill
\begin{minipage}{0.49\linewidth}\label{fig5b}
  \centerline{\includegraphics[width=2in]{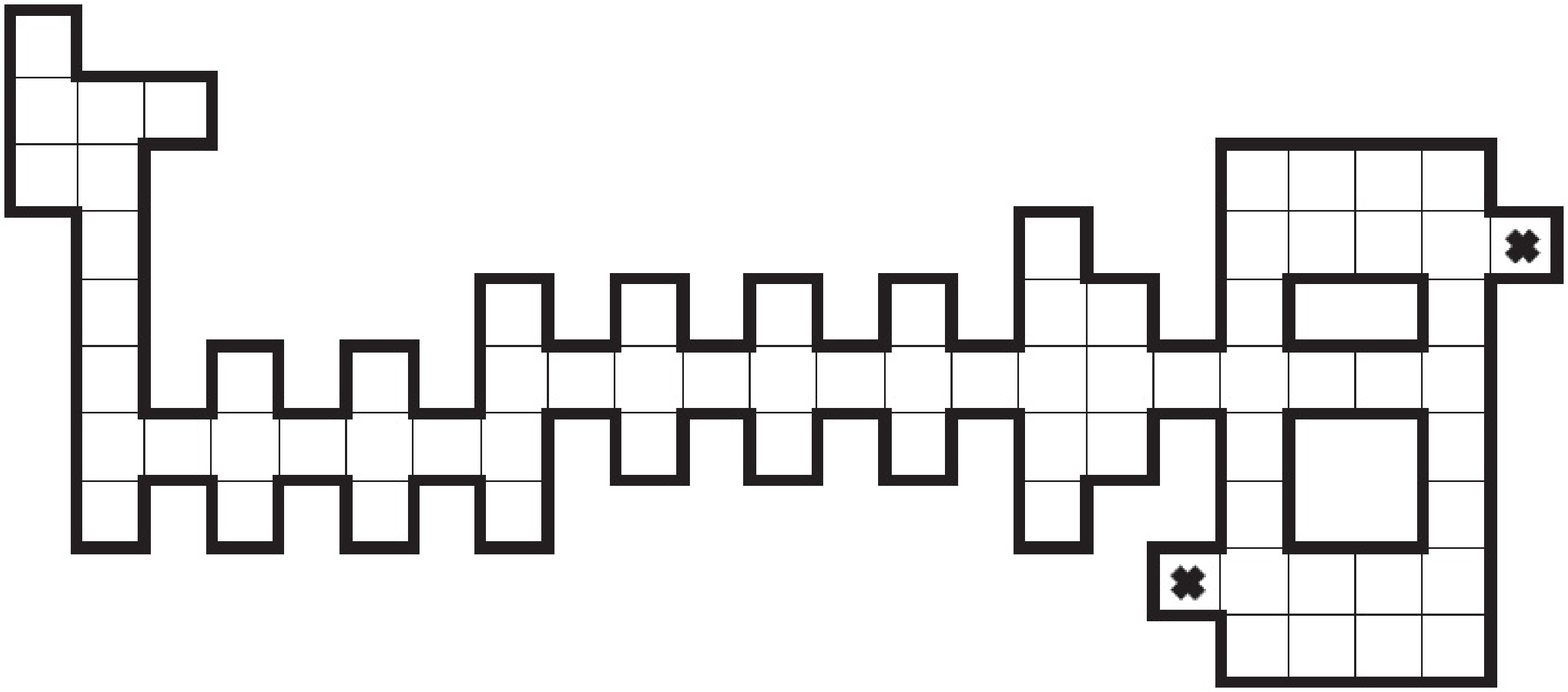}}
  \vspace{9pt}
  \centerline{(b) SUNY}
\end{minipage}
\vfill
\begin{minipage}{0.49\linewidth}\label{fig5c}
  \centerline{\includegraphics[width=2in]{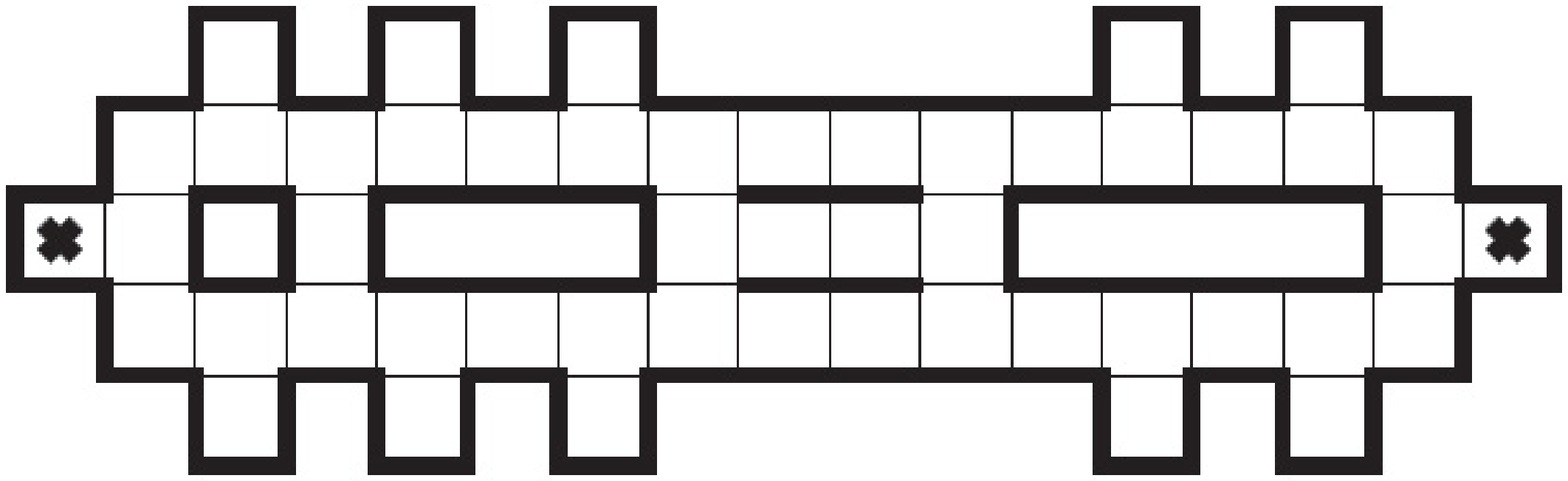}}
  \vspace{15pt}
  \centerline{(c) MIT}
\end{minipage}
\hfill
\begin{minipage}{0.49\linewidth}\label{fig5d}
  \centerline{\includegraphics[width=1.1in]{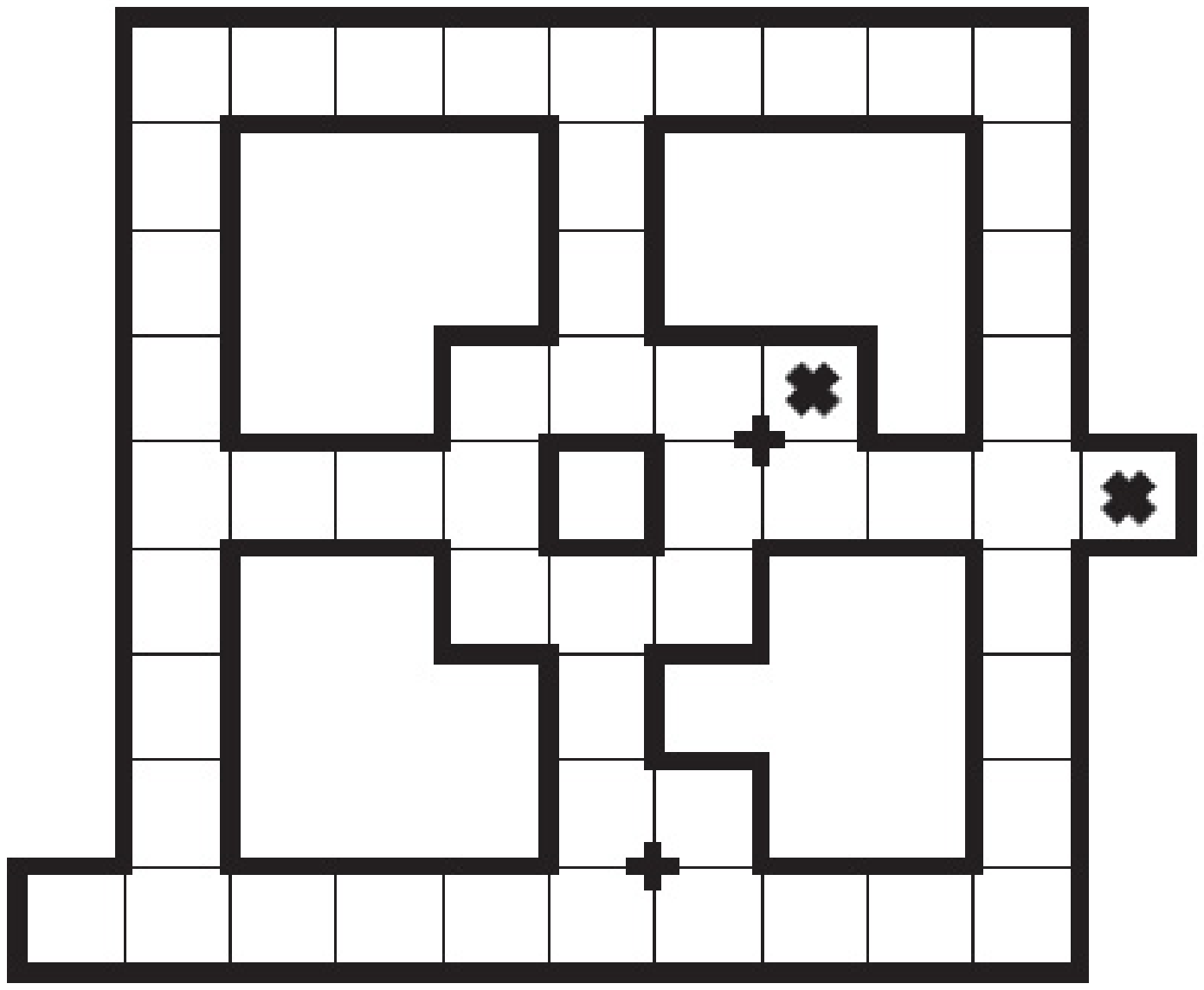}}
  \vspace{-3pt}
  \centerline{(d) PENTAGON}
\end{minipage}
\vfill
\vspace{5pt}
\begin{minipage}{0.49\linewidth}\label{fig5e}
  \centerline{\includegraphics[width=1.4in]{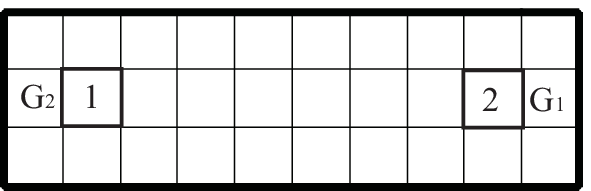}}
  \vspace{15pt}
  \centerline{(e) GW\_{nju}}
\end{minipage}
\hfill
\begin{minipage}{0.49\linewidth}\label{fig5f}
  \centerline{\includegraphics[width=0.7in]{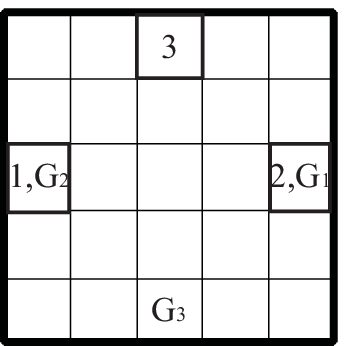}}
  \vspace{-1pt}
  \centerline{(f) GWa3}
\end{minipage}

\caption{Grid world games.} \label{fig5}
\end{figure}

\begin{table*}[!bp]
\centering
\small
\caption{The learning steps of the final episode of each tested benchmark map.}\label{Table1}
\renewcommand\arraystretch{1.5}
\begin{tabular}{|c|c|c|c|c|c|c|}
\hline 
    $$ & $ISR$ & $SUNY$ & $MIT$ &   $PENTAGON$ &    $GW\_{nju}$ & $GWa3$\\
    \hline
    $CQ-learning$ & $8.91$ &    $10.70$ &   $19.81$ &   $15.32$ &   $12.65$ &   $8.65$\\
    \hline
    $ILVFT$ &   $13.11$ &   $\textbf{10.38}$ &  $\textbf{18.67}$ &  $14.18$ &   $\textbf{10.94}$ &  $11.40$\\
    \hline
    $NegoQVFT$ &    $8.36$ &    $12.98$ &   $19.81$ &   $\textbf{8.55}$ &   $10.95$ &   $\textbf{8.31}$\\
    \hline
    $NegoSI$ &  $\textbf{7.48}$ &   $10.92$ &   $21.29$ &   $10.30$ &   $12.11$ &   $8.87$\\
    \hline
    $The \ optimal \ policy$ &  $6$ &   $10$ &  $18$ &  $8$ &   $10$ &  $8$\\
    \hline
    \end{tabular}
\end{table*}

The proposed NegoSI algorithm is evaluated in the grid world games
presented by Melo and Veloso \cite{Melo and Veloso 2009}, which
are shown in Fig. \ref{fig5}. The first four benchmarks, i.e.,
ISR, SUNY, MIT and PENTAGON (shown as Fig. \ref{fig5}(a)-(d),
respectively), are two-agent games, where the cross symbols denote
the initial locations of each agent and the goals of the other
agent. In addition, we design two highly competitive games to
further test the algorithms, namely, GW\_{nju} (Fig.
\ref{fig5}(e)) and GWa3 (Fig. \ref{fig5}(f)). The game GW\_{nju}
has two agents and the game GWa3 has three agents. The initial
location and the goal of agent $i$ are represented by the number
$i$ and $G_i$, respectively.

\begin{figure}[!tp]
\begin{minipage}{0.49\linewidth}\label{fig6a}
  \centerline{\includegraphics[width=1.8in]{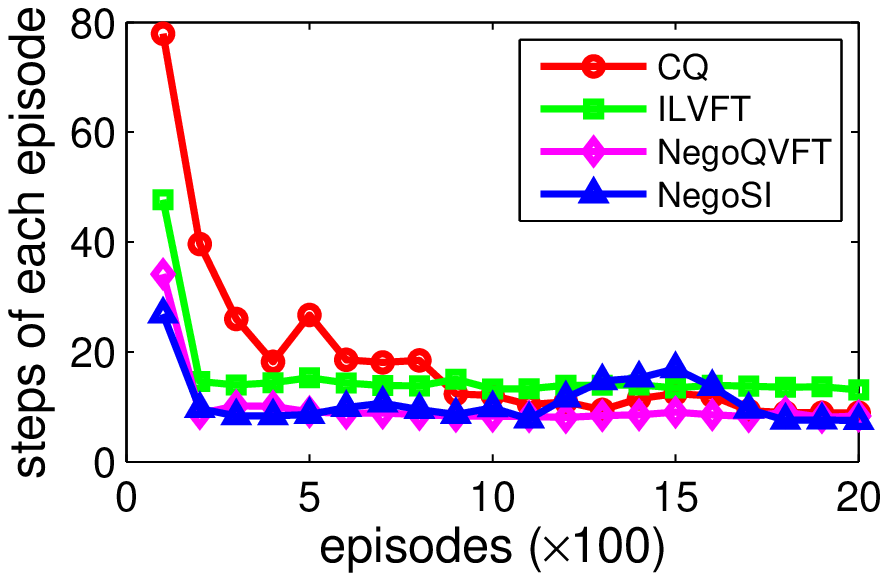}}
  \centerline{(a) ISR}
\end{minipage}
\hfill
\begin{minipage}{0.49\linewidth}\label{fig6b}
  \centerline{\includegraphics[width=1.8in]{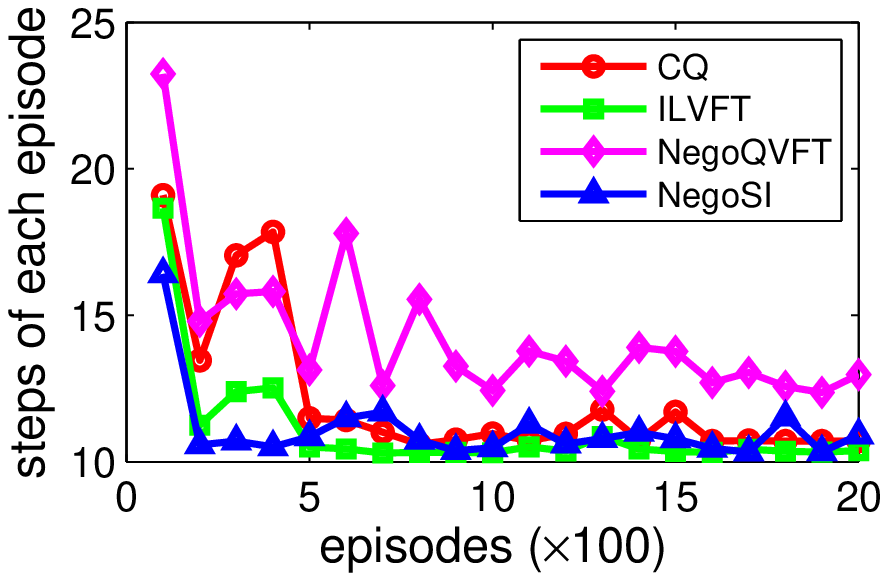}}
  \centerline{(b) SUNY}
\end{minipage}
\vfill
\begin{minipage}{0.49\linewidth}\label{fig6c}
  \centerline{\includegraphics[width=1.8in]{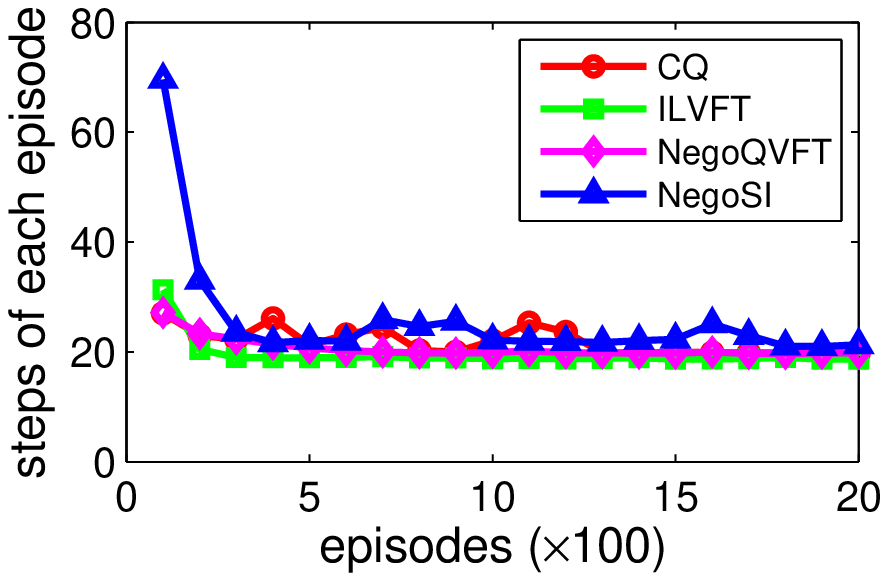}}
  \centerline{(c) MIT}
\end{minipage}
\hfill
\begin{minipage}{0.49\linewidth}\label{fig6d}
  \centerline{\includegraphics[width=1.8in]{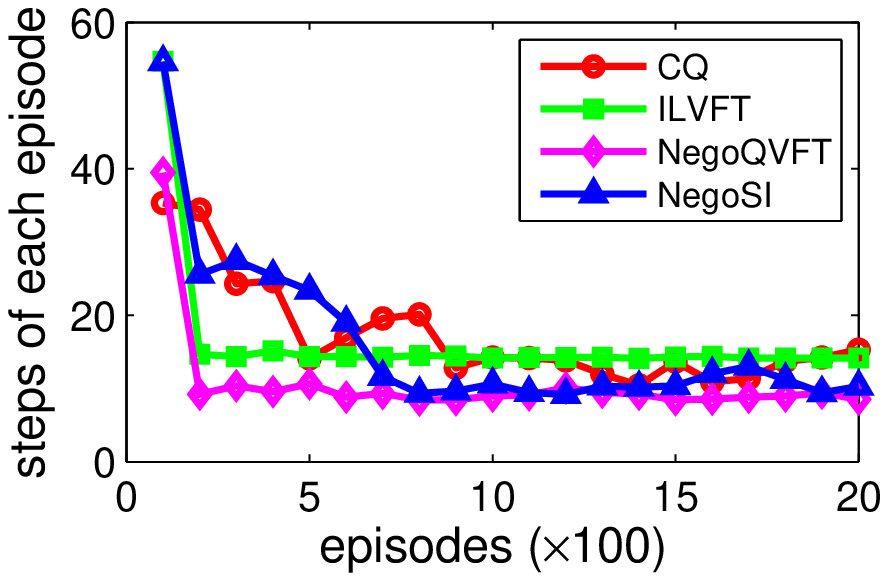}}
  \centerline{(d) PENTAGON}
\end{minipage}
\vfill
\begin{minipage}{0.49\linewidth}\label{fig6e}
  \centerline{\includegraphics[width=1.8in]{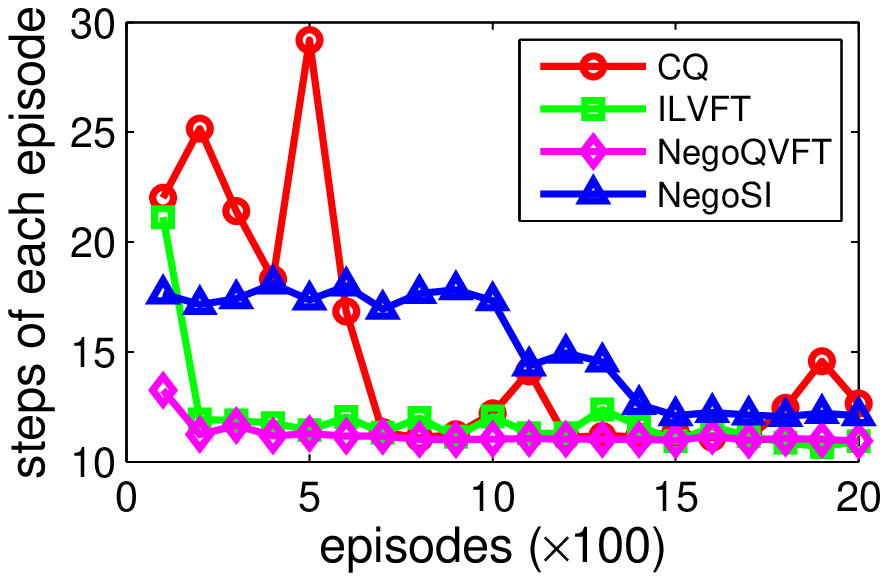}}
  \centerline{(e) GW\_{nju}}
\end{minipage}
\hfill
\begin{minipage}{0.49\linewidth}\label{fig6f}
  \centerline{\includegraphics[width=1.8in]{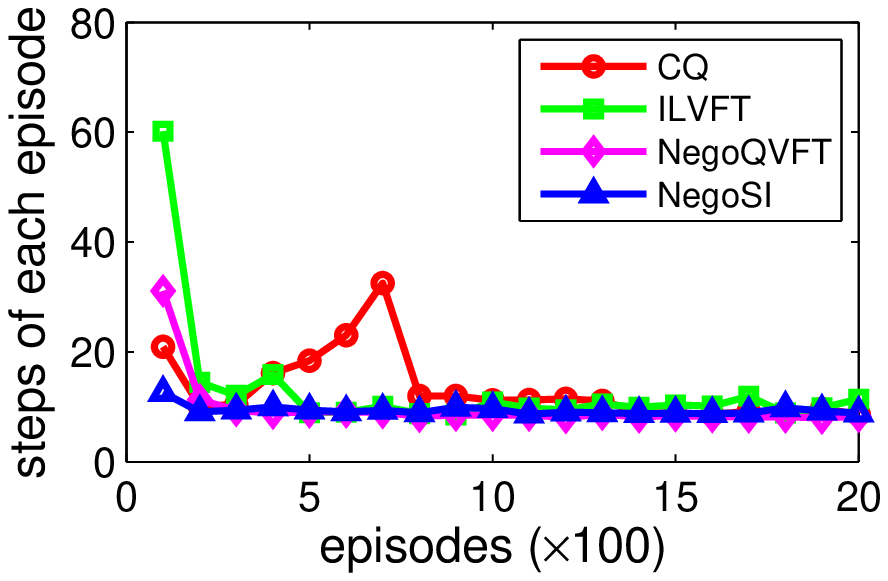}}
  \centerline{(f) GWa3}
\end{minipage}

\caption{SEE (steps of each episode) for each tested benchmark map.}
\label{fig6}
\end{figure}

We first examine the performances of the tested algorithms
regarding the SEE (steps of each episode) for each benchmark map
(See Fig. \ref{fig6}). The state-of-the-art value function
transfer mechanism helps NegoQ and ILs to converge fast in all
games except for SUNY. NegoSI also has good convergence
characteristics. CQ-learning, however, has a less stable learning
curve especially in PENTAGON, GW\_{nju} and GWa3. This is
reasonable since in these highly competitive games, the agents'
prior-learned optimal policies always have conflicts and
CQ-learning cannot find the equilibrium joint action to coordinate
them. When more collisions occur, the agents are frequently
bounced to previous states and forced to take more steps before
reaching their goals. The learning steps of the final episode for
each algorithm in each benchmark map are shown in Table
\ref{Table1}. In ISR, NegoSI converges to 7.48, which is the
closest one to the value obtained by the optimal policy. In other
cases, NegoSI achieves the learning step of the final episode
between 105.1\% and 120.4\% to the best solution.

\begin{figure*}[tp]
\begin{minipage}{0.2\linewidth}\label{fig7a}
  \centerline{\includegraphics[width=1.8in]{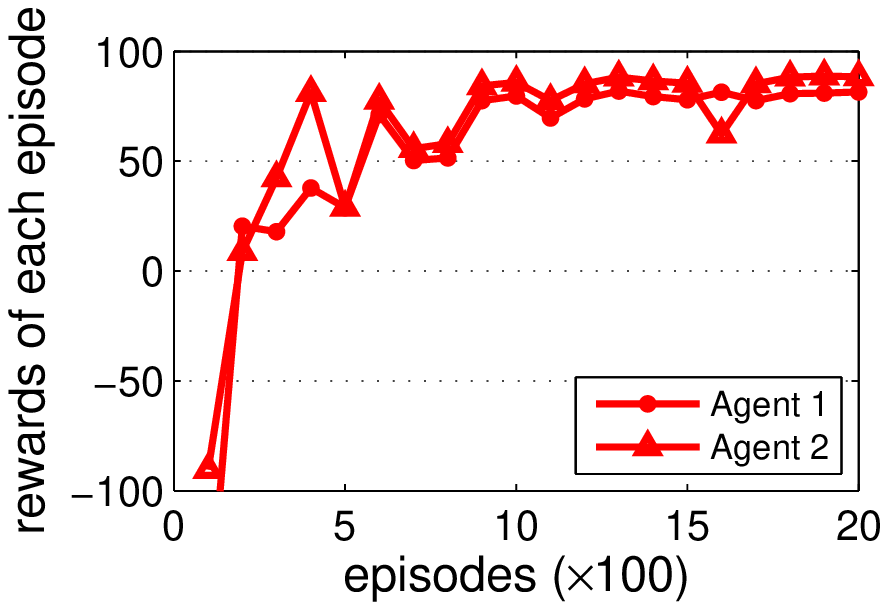}}
  \centerline{(a) CQ-learning}
\end{minipage}
\hfill
\begin{minipage}{0.2\linewidth}\label{fig7b}
  \centerline{\includegraphics[width=1.8in]{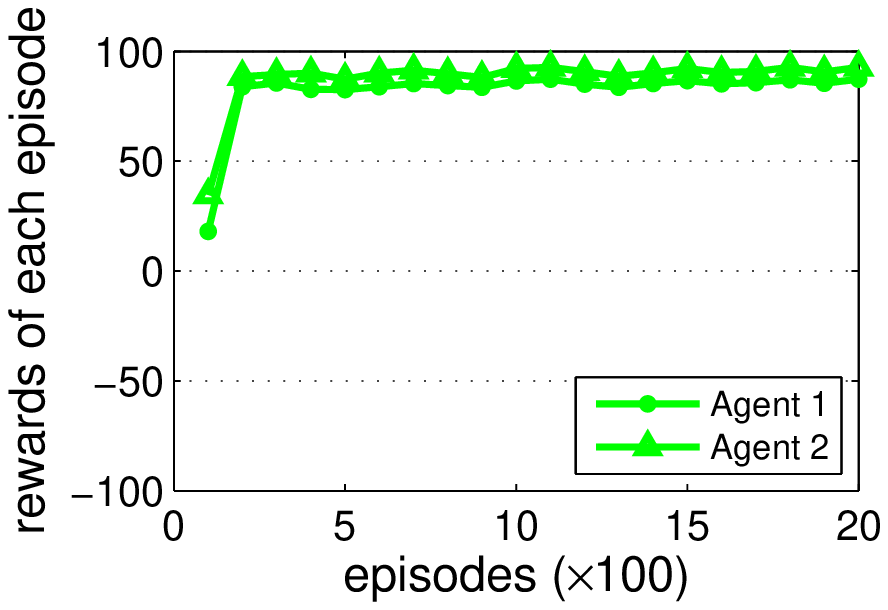}}
  \centerline{(b) ILVFT}
\end{minipage}
\hfill
\begin{minipage}{0.2\linewidth}\label{fig7c}
  \centerline{\includegraphics[width=1.8in]{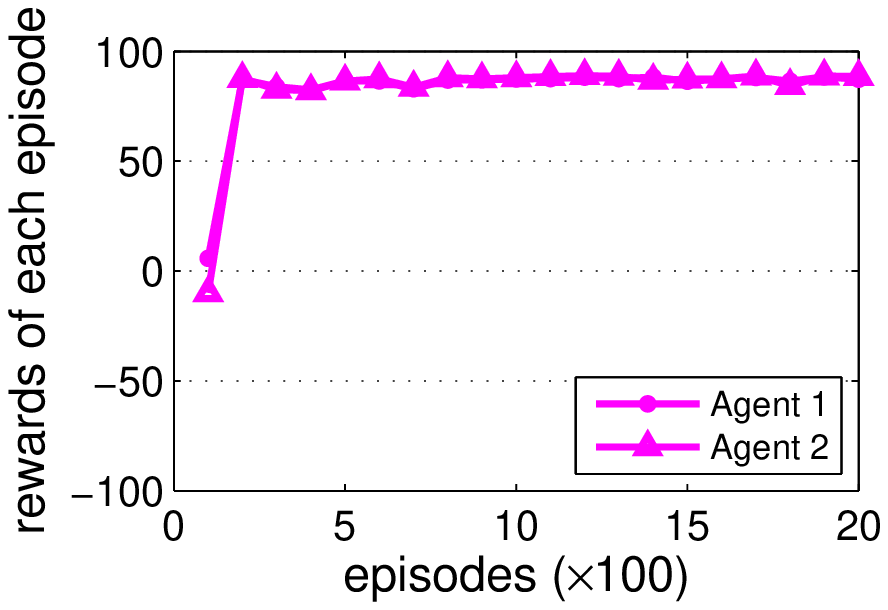}}
  \centerline{(c) NegoQVFT}
\end{minipage}
\hfill
\begin{minipage}{0.2\linewidth}\label{fig7d}
  \centerline{\includegraphics[width=1.8in]{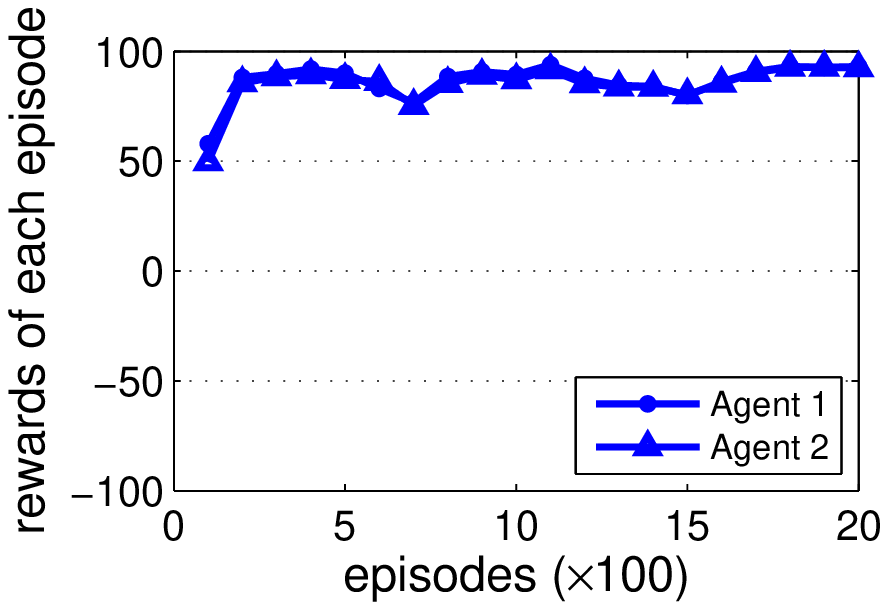}}
  \centerline{(d) NegoSI}
\end{minipage}
\caption{REE (rewards of each episode) for each tested algorithm in ISR.}
\label{fig7}
\vspace{-6mm}
\end{figure*}

\begin{figure*}[tp]
\begin{minipage}{0.2\linewidth}\label{fig8a}
  \centerline{\includegraphics[width=1.8in]{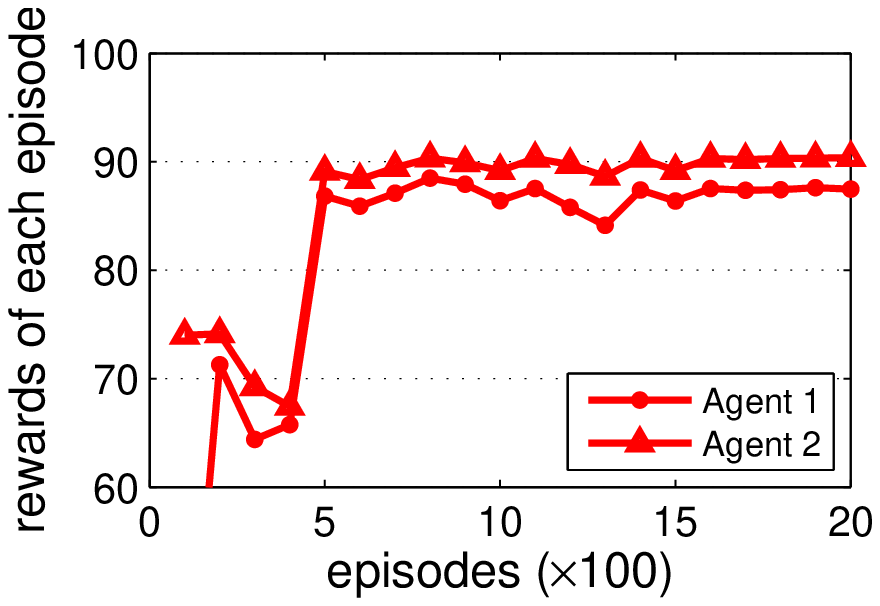}}
  \centerline{(a) CQ-learning}
\end{minipage}
\hfill
\begin{minipage}{0.2\linewidth}\label{fig8b}
  \centerline{\includegraphics[width=1.8in]{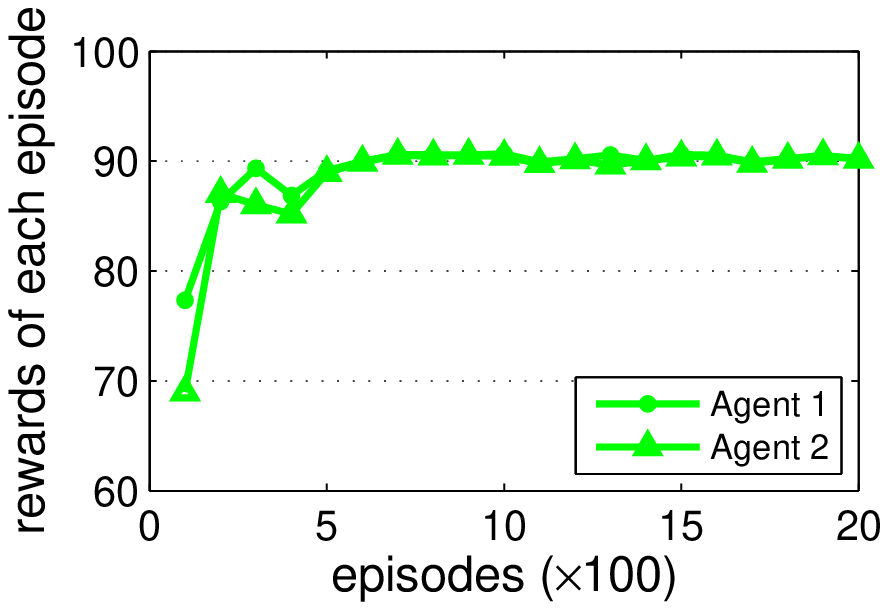}}
  \centerline{(b) ILVFT}
\end{minipage}
\hfill
\begin{minipage}{0.2\linewidth}\label{fig8c}
  \centerline{\includegraphics[width=1.8in]{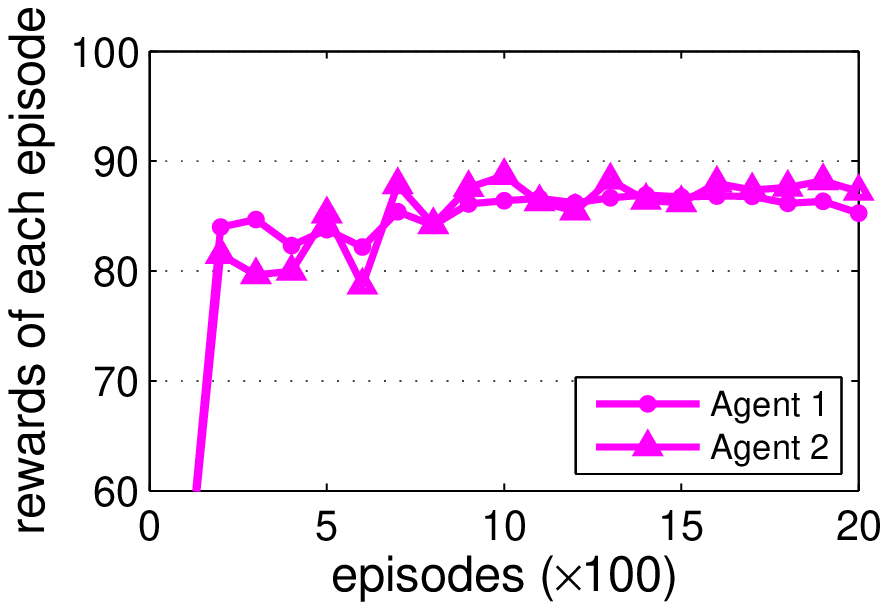}}
  \centerline{(c) NegoQVFT}
\end{minipage}
\hfill
\begin{minipage}{0.2\linewidth}\label{fig8d}
  \centerline{\includegraphics[width=1.8in]{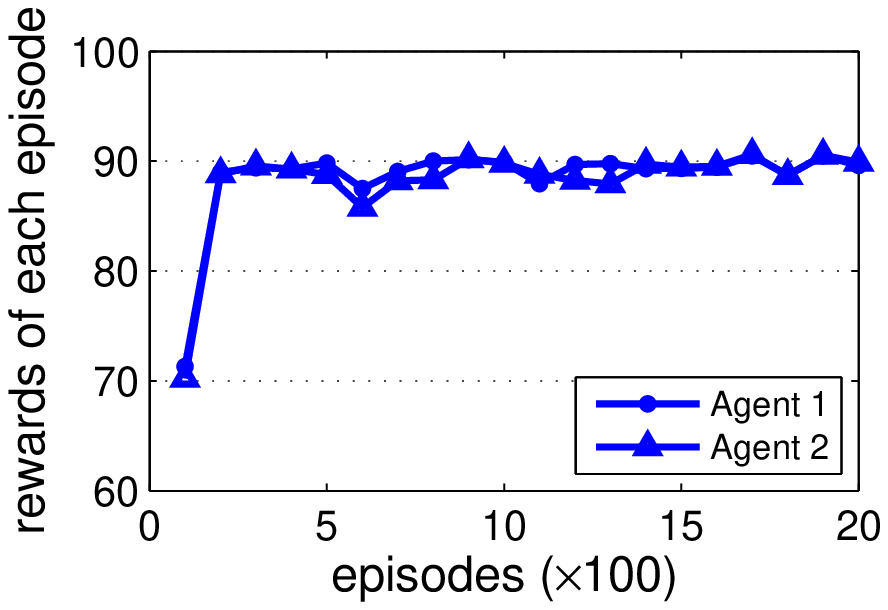}}
  \centerline{(d) NegoSI}
\end{minipage}
\caption{REE (rewards of each episode) for each tested algorithm in SUNY.}
\label{fig8}
\vspace{-6mm}
\end{figure*}

\begin{figure*}[tp]
\begin{minipage}{0.2\linewidth}\label{fig9a}
  \centerline{\includegraphics[width=1.8in]{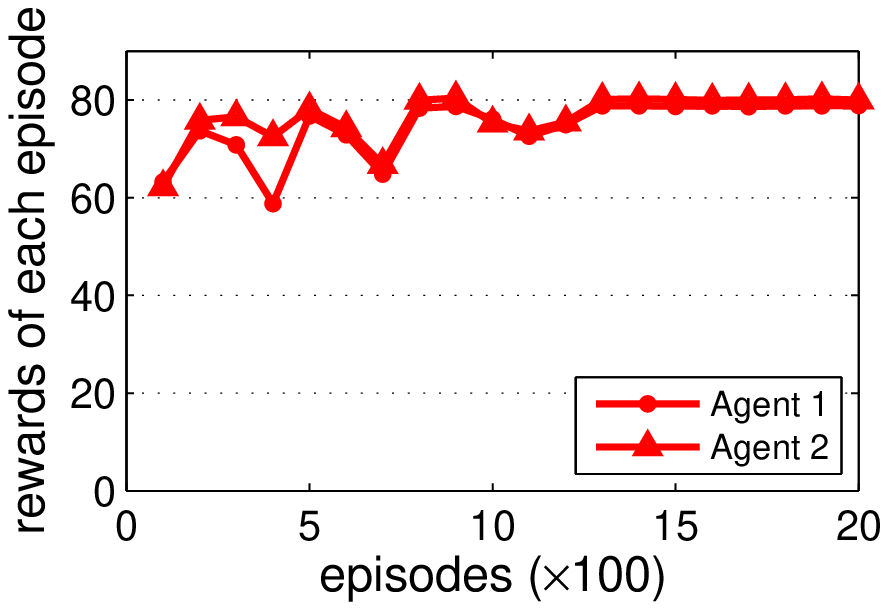}}
  \centerline{(a) CQ-learning}
\end{minipage}
\hfill
\begin{minipage}{0.2\linewidth}\label{fig9b}
  \centerline{\includegraphics[width=1.8in]{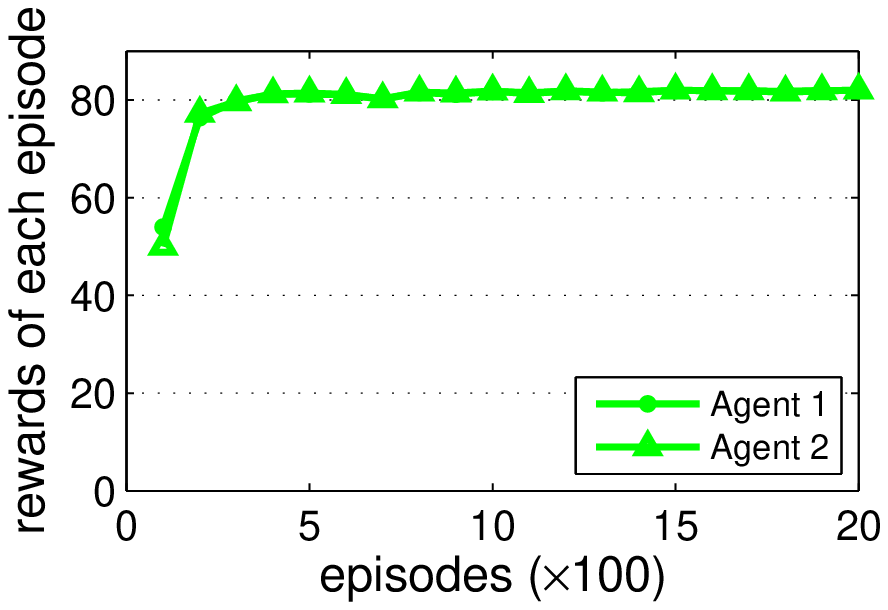}}
  \centerline{(b) ILVFT}
\end{minipage}
\hfill
\begin{minipage}{0.2\linewidth}\label{fig9c}
  \centerline{\includegraphics[width=1.8in]{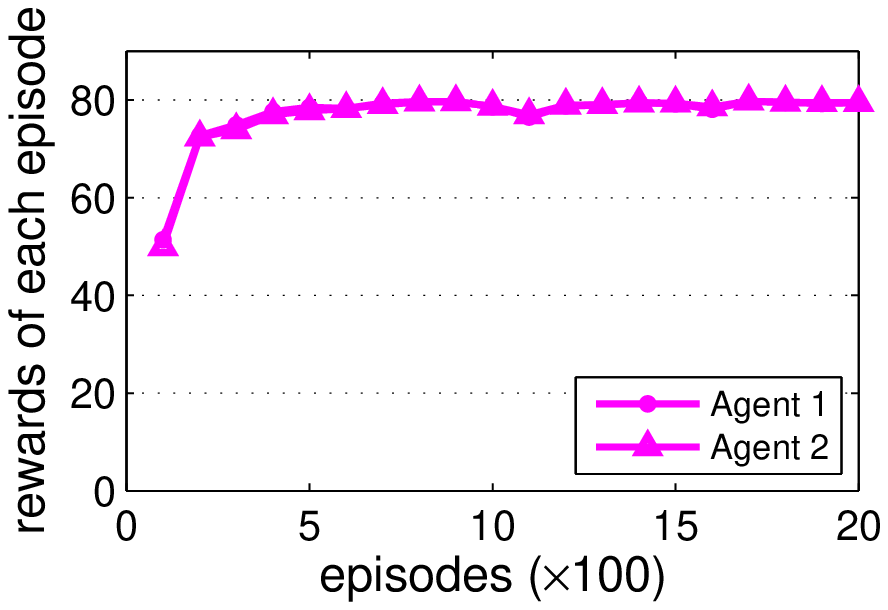}}
  \centerline{(c) NegoQVFT}
\end{minipage}
\hfill
\begin{minipage}{0.2\linewidth}\label{fig9d}
  \centerline{\includegraphics[width=1.8in]{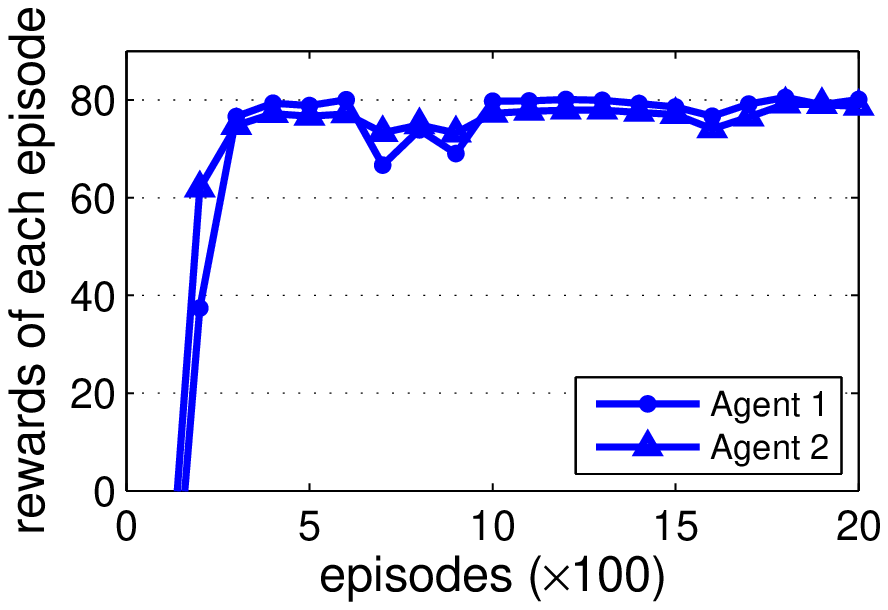}}
  \centerline{(d) NegoSI}
\end{minipage}
\caption{REE (rewards of each episode) for each tested algorithm in MIT.}
\label{fig9}
\vspace{-6mm}
\end{figure*}

\begin{figure*}[tp]
\begin{minipage}{0.2\linewidth}\label{fig10a}
  \centerline{\includegraphics[width=1.8in]{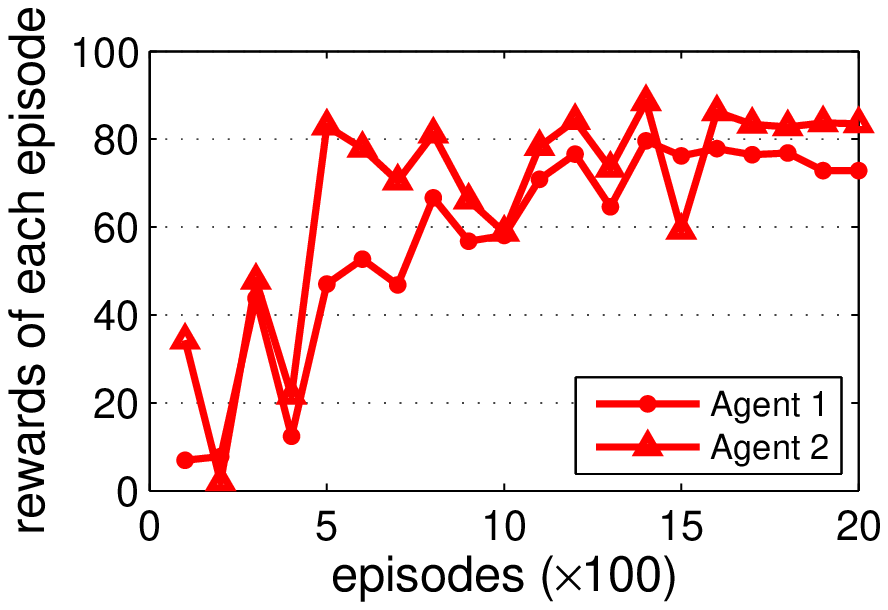}}
  \centerline{(a) CQ-learning}
\end{minipage}
\hfill
\begin{minipage}{0.2\linewidth}\label{fig10b}
  \centerline{\includegraphics[width=1.8in]{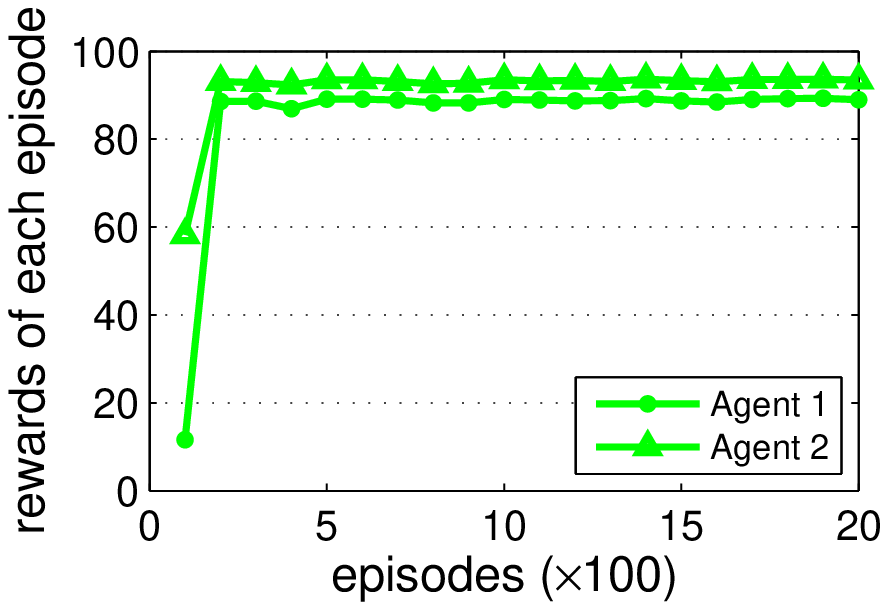}}
  \centerline{(b) ILVFT}
\end{minipage}
\hfill
\begin{minipage}{0.2\linewidth}\label{fig10c}
  \centerline{\includegraphics[width=1.8in]{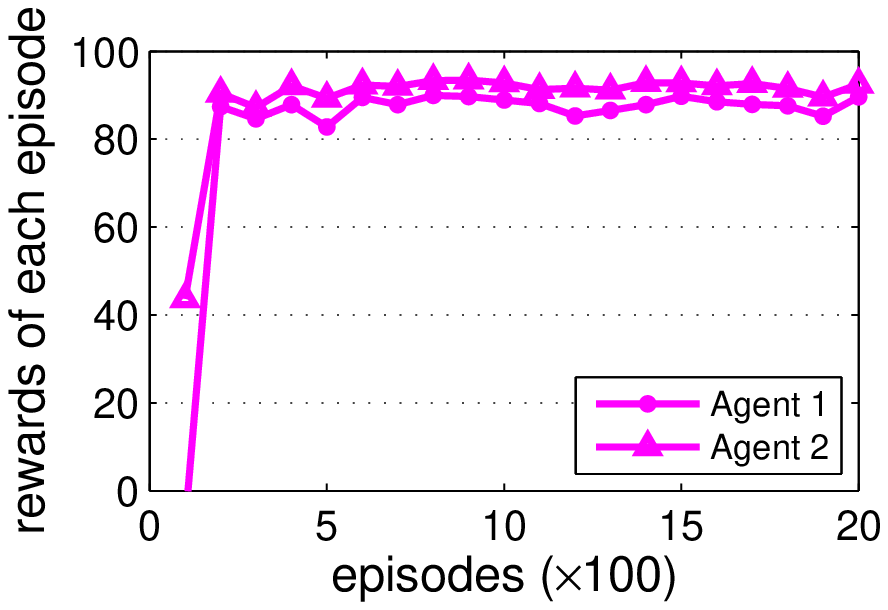}}
  \centerline{(c) NegoQVFT}
\end{minipage}
\hfill
\begin{minipage}{0.2\linewidth}\label{fig10d}
  \centerline{\includegraphics[width=1.8in]{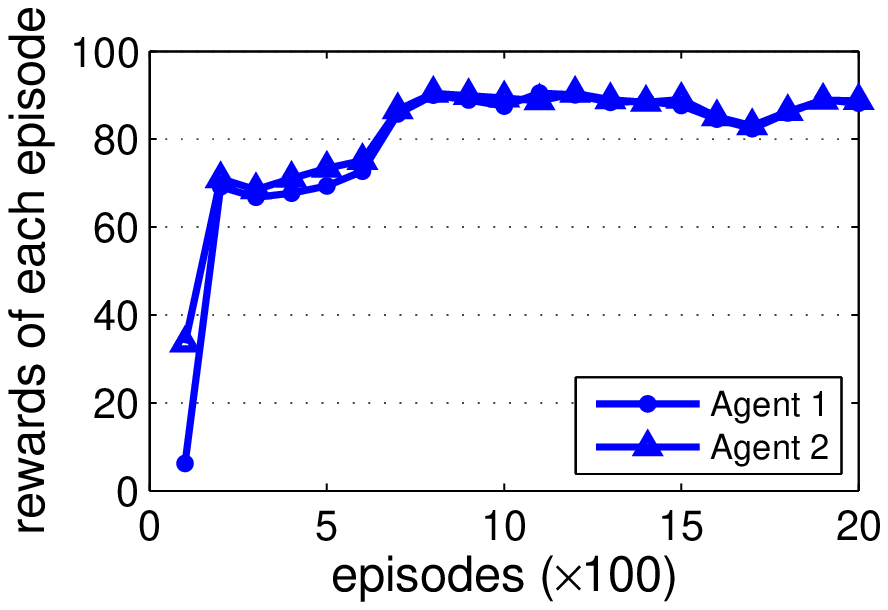}}
  \centerline{(d) NegoSI}
\end{minipage}
\caption{REE (rewards of each episode) for each tested algorithm in PENTAGON.}
\label{fig10}
\vspace{-6mm}
\end{figure*}

\begin{figure*}[tp]
\begin{minipage}{0.2\linewidth}\label{fig11a}
  \centerline{\includegraphics[width=1.8in]{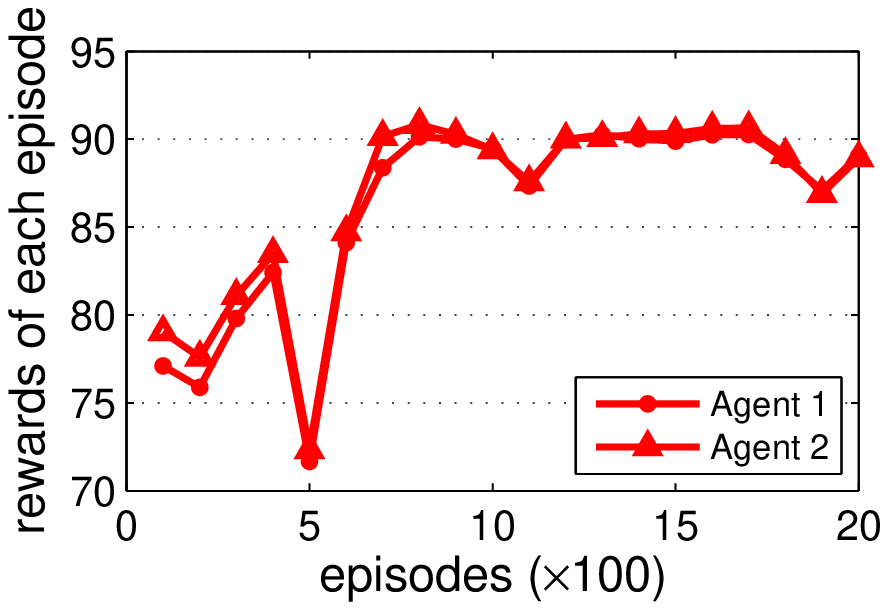}}
  \centerline{(a) CQ-learning}
\end{minipage}
\hfill
\begin{minipage}{0.2\linewidth}\label{fig11b}
  \centerline{\includegraphics[width=1.8in]{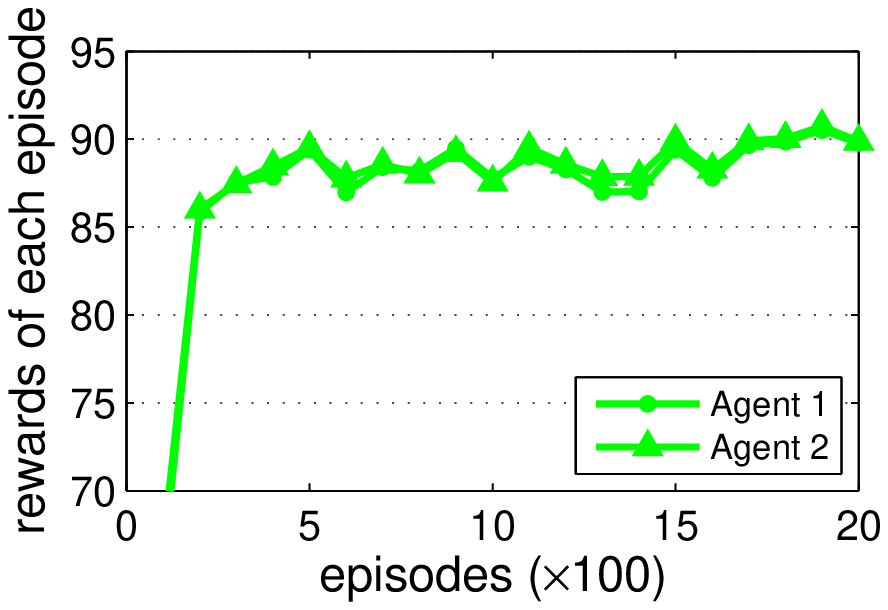}}
  \centerline{(b) ILVFT}
\end{minipage}
\hfill
\begin{minipage}{0.2\linewidth}\label{fig11c}
  \centerline{\includegraphics[width=1.8in]{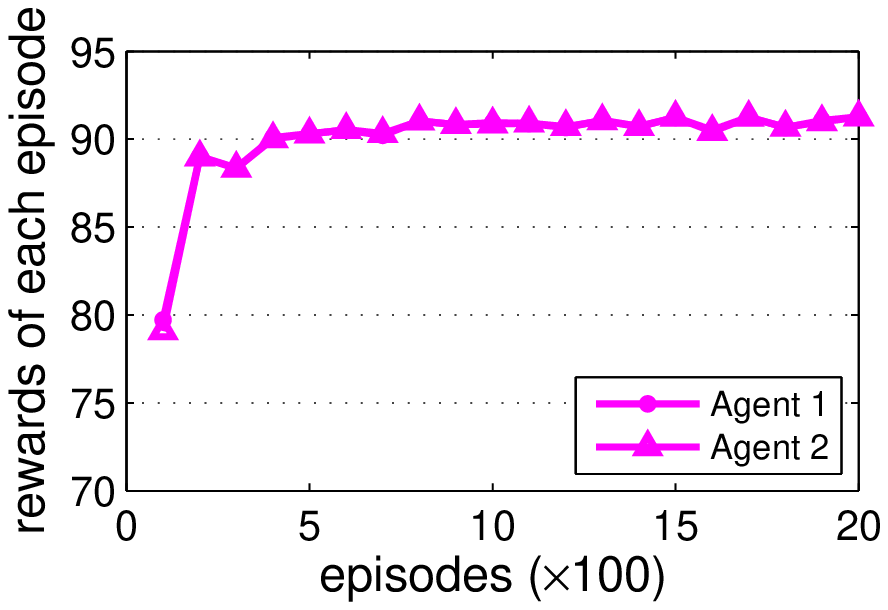}}
  \centerline{(c) NegoQVFT}
\end{minipage}
\hfill
\begin{minipage}{0.2\linewidth}\label{fig11d}
  \centerline{\includegraphics[width=1.8in]{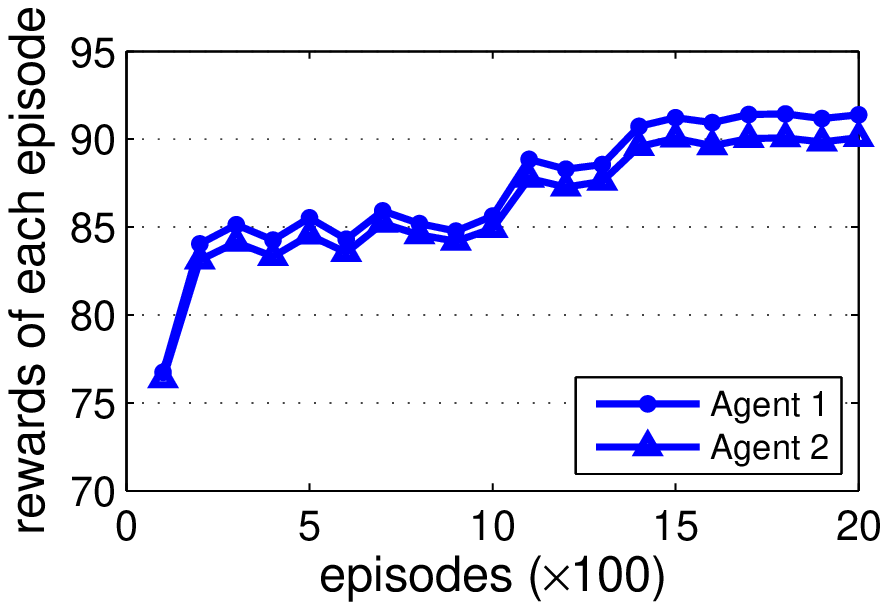}}
  \centerline{(d) NegoSI}
\end{minipage}
\caption{REE (rewards of each episode) for each tested algorithm in GW\_{nju}.}
\label{fig11}
\vspace{-6mm}
\end{figure*}

\begin{figure*}[tp]
\begin{minipage}{0.2\linewidth}\label{fig12a}
  \centerline{\includegraphics[width=1.8in]{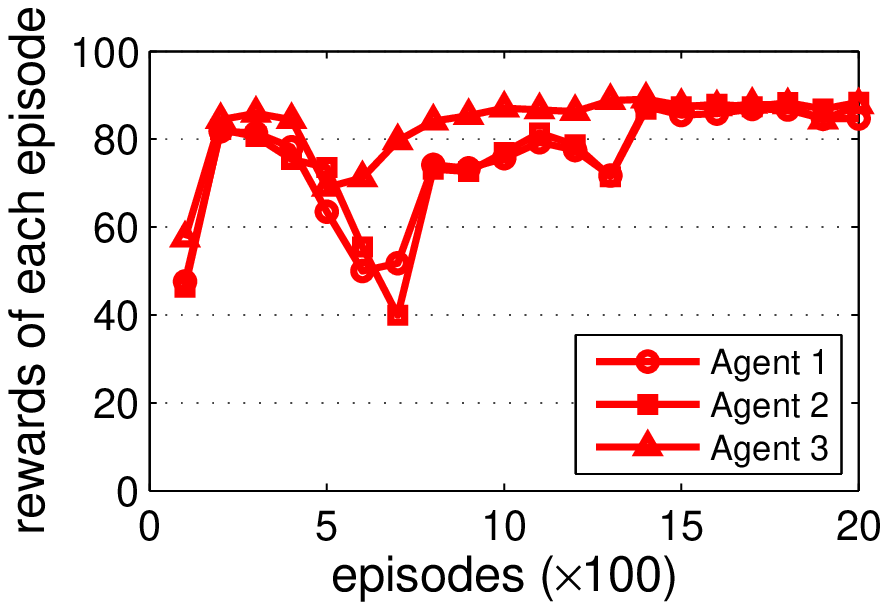}}
  \centerline{(a) CQ-learning}
\end{minipage}
\hfill
\begin{minipage}{0.2\linewidth}\label{fig12b}
  \centerline{\includegraphics[width=1.8in]{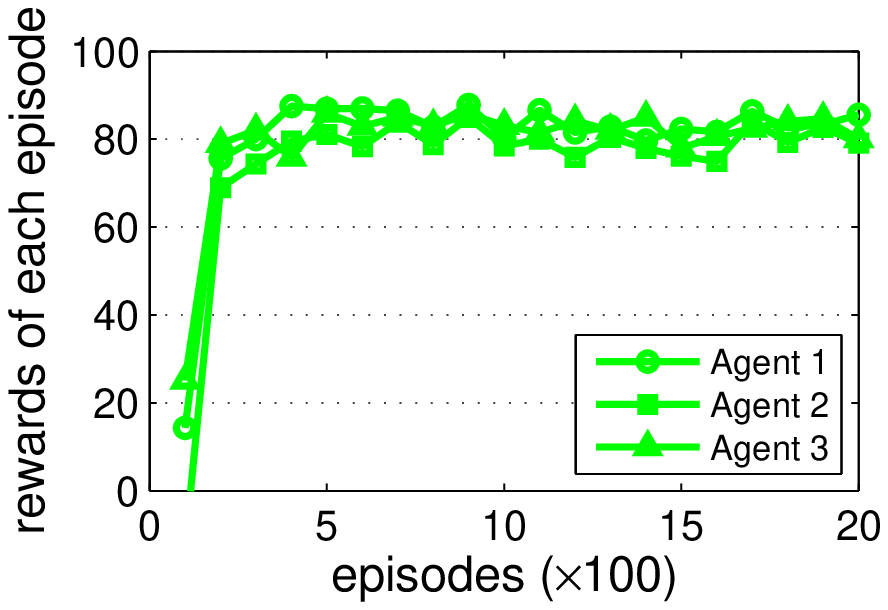}}
  \centerline{(b) ILVFT}
\end{minipage}
\hfill
\begin{minipage}{0.2\linewidth}\label{fig12c}
  \centerline{\includegraphics[width=1.8in]{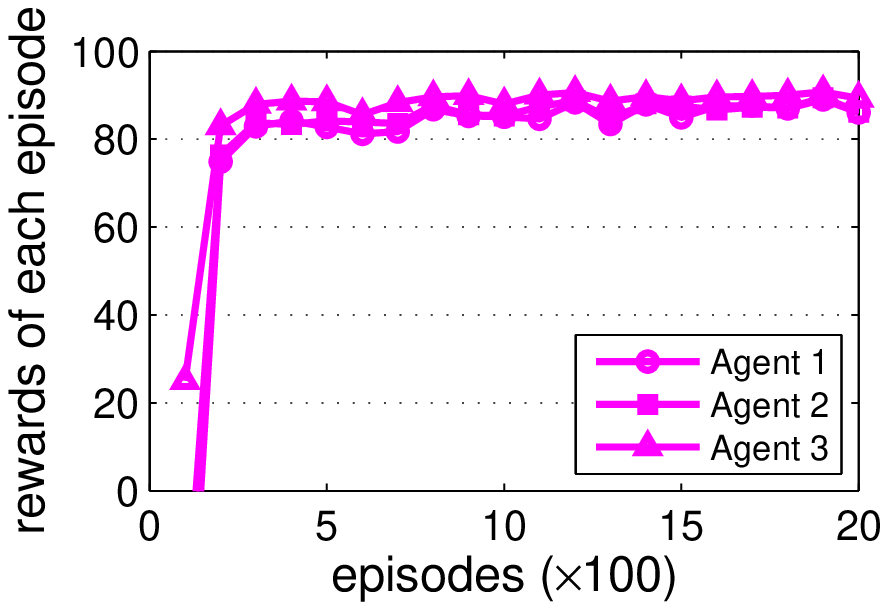}}
  \centerline{(c) NegoQVFT}
\end{minipage}
\hfill
\begin{minipage}{0.2\linewidth}\label{fig12d}
  \centerline{\includegraphics[width=1.8in]{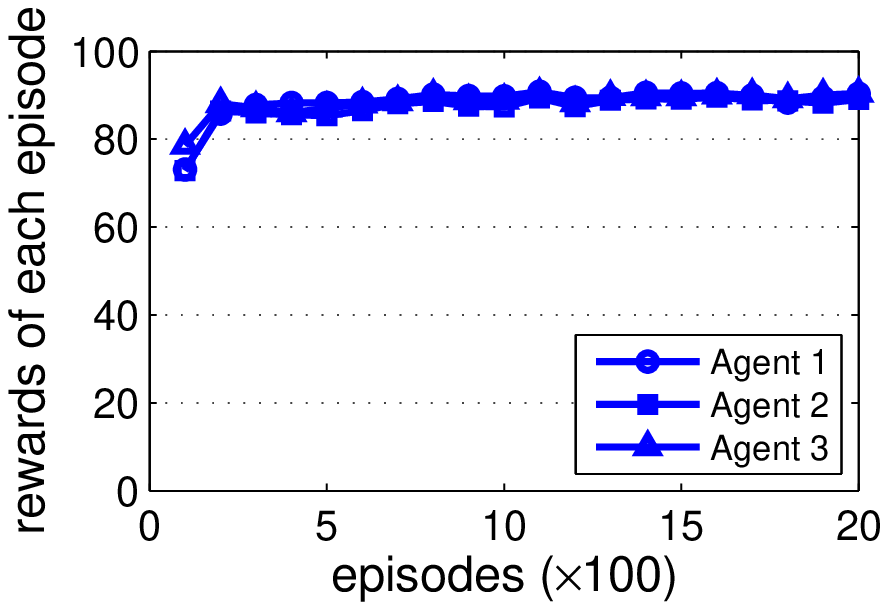}}
  \centerline{(d) NegoSI}
\end{minipage}
\caption{REE (rewards of each episode) for each tested algorithm in GWa3.}
\label{fig12}
\vspace{-6mm}
\end{figure*}

Then we analyze the REE (rewards of each episode) criterion of
these tested algorithms (See Fig. \ref{fig7}-\ref{fig12}). The
results vary for each map. In ISR, NegoSI generally achieves the
highest REE through the whole learning process, which shows fewer
collisions and more cooperations between the learning agents.
Also, the difference between the two agents' reward values is
small, which reflects the fairness of the learning process of
NegoSI. The agents are more independent in SUNY. Each of them has
three candidates for the single-agent optimal policy. In this
setting, ILVFT has good performance. NegoSI shows its fairness and
high REE value compared with NegoVFT and CQ-learning. The agents
in MIT have more choices of collision avoidances. All the tested
algorithms obtain the final reward of around 80. However, the
learning curves of CQ-learning are relatively unstable. In
PENTAGON, NegoSI proves its fairness and achieves as good
performance as algorithms do with the value function transfer.

Other than these above benchmarks, we give two highly competitive
games to test the proposed algorithm. In both games, the agents'
single-agent optimal policies conflict with each other and need
appropriate coordination. In GW\_{nju}, the learning curves of
NegoSI finally converge to 91.39 and 90.09, which are very close
to the optimal final REE values as 93 and 91, respectively.
Similar to NegoSI, NegoVFT achieves the final REE values of 91.27
and 91.25. However, NegoSI can negotiate for a fixed and safer
policy that allows one agent to always move greedily and the other
to avoid collision (while NegoVFT cannot), which shows the better
coordination ability of NegoSI. For the three-agent grid world
GWa3, even though NegoSI does not have the lowest SEE, it achieves
the highest REE through the whole learning process. For one thing, this result demonstrates the scalability
of NegoSI in the three-agent setting. For another, it shows that the agents using NegoSI have the ability to avoid
collisions to obtain more rewards while the agents using traditional MARL methods have less desire to coordinate and
therefore lose their rewards. Actually, even though we developed our method with non-cooperative multi-agent model, it does not necessarily mean that the agents are egoistic. Thanks to the negotiation mechanism, agents learn to benefit themselves while doing little harm to others, which shows an evidence of cooperation.

\begin{table*}[!tp]
\centering
\small
\caption{Average runtime for each tested map.}\label{Table2}
\renewcommand\arraystretch{1.5}
\begin{tabular}{|c|c|c|c|c|c|c|}
\hline 
  \centering
    $$ & $ISR$ & $SUNY$ & $MIT$ &   $PENTAGON$ &    $GW\_{nju}$ & $GWa3$\\
    \hline
    $CQ-learning$ & $8.54$ &    $5.91$  & $13.68$   & $8.52$ &  $5.07$  & $5.95$\\
    \hline
    $ILVFT$ &   $4.91$ &    $4.14$  & $9.78$ & $6.56$ & $2.53$  & $7.21$\\
    \hline
    $NegoQVFT$ &    $13.92$ &   $20.18$ &   $36.84$ & $18.45$ & $21.89$ &     $50.48$\\
    \hline
    $NegoSI$ &  $16.74$ &   $7.33$  & $19.58$   & $16.18$ & $7.08$  & $16.41$\\
    \hline
    $QL \ for \ Single \ Policy$ & $0.51$  & $0.45$ &  $1.29$  & $0.05$    & $0.19$ &  $0.07$\\
    \hline
    \end{tabular}
\end{table*}

The results regarding AR (average runtime) are shown in Table \ref{Table2}. ILVFT has the fastest learning speed, which is only five to ten times slower than Q-learning to learn single policy. CQ-learning only considers joint states in ``coordination state" and it also has a relatively small computational complexity. The AR of NegoVFT is five to ten times more than that of ILVFT. This is reasonable since NegoVFT learns in the whole joint state-action space and computes the equilibrium for each joint state. The learning speed of NegoSI is slower than CQ-learning but faster than NegoVFT. Even if NegoSI adopts the sparse-interaction based learning framework and has computational complexity similar to CQ-learning, it needs to search for the equilibrium joint action in the ``coordination state", which slows down the learning process.

\subsection{A real-world application: intelligent warehouse systems}\label{Sec4.2}

\begin{figure}[!bp]
  \centering
  \includegraphics[width=1.9in]{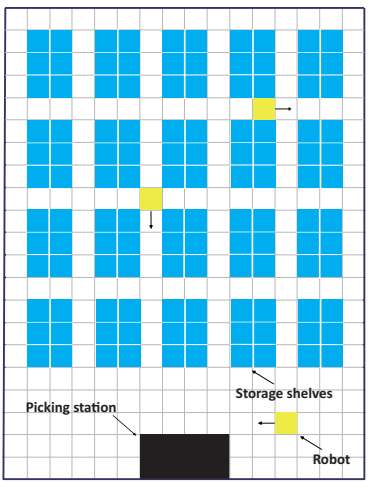}\\
  \caption{Simulation platform of the intelligent warehouse system.}\label{fig14}
\end{figure}

MARL has been widely used in such simulated domains as grid worlds \cite{Busoniu and Babuska 2008}, but few applications have been found for realistic problems comparing to single-agent RL algorithms \cite{Liu and Yang 2015}-\cite{Valasek and Doebbler 2008} or MAS algorithms \cite{Sabattini and Secchi 2015}-\cite{Zhang and Feng 2015}. In this paper, we apply the proposed NegoSI algorithm to an intelligent warehouse problem. Previously, single agent path planning methods have been successfully used in complex systems \cite{Zhou and Shi 2014} \cite{Viet and Kyaw 2011}, however, the intelligent warehouse employs a team of mobile robots to transport objects and single-agent path planning methods frequently cause collisions \cite{Wurman and D'Andrea 2008}. So we solve the multi-agent coordination problem in a learning way.

The intelligent warehouse is made up of three parts: picking stations, robots and storage shelves (shown as in Fig. \ref{fig14}). There are generally four steps of the order fulfillment process for intelligent warehouse systems:

\begin{enumerate}
  \item Input and decomposition: input and decompose orders into separated tasks;
  \item Task allocation: the central controller allocates the tasks to corresponding robots using task allocation algorithms (e.g., the Auction method);
  \item Path planning: robots plan their transportation paths with a single-agent path planning algorithm;
  \item Transportation and collision avoidance: robots transport their target storage shelves to the picking station and then bring them back to their initial positions. During the transportation process, robots use sensors to detect shelves and other robots to avoid collisions.
\end{enumerate}

We focus on solving the multi-robot path planning and collision avoidance in a MARL way and ignore the first two steps of the order fulfillment process by requiring each robot to finish a certain number of random tasks. The simulation platform of an intelligent warehouse system is shown as in Fig. \ref{fig14}, which is a $16\times 21$ grid environment and each grid is of size $0.5m\times0.5m$ in real world. Shaded grids are storage shelves which cannot be passed through. The state space of each robot is made up of two parts: the location and the task number. Each robot has 4 actions, namely, ``up", ``down", ``left" and ``right".

\begin{figure*}[!bp]
\begin{minipage}{0.49\linewidth}\label{fig15a}
  \centerline{\includegraphics[width=2.6in]{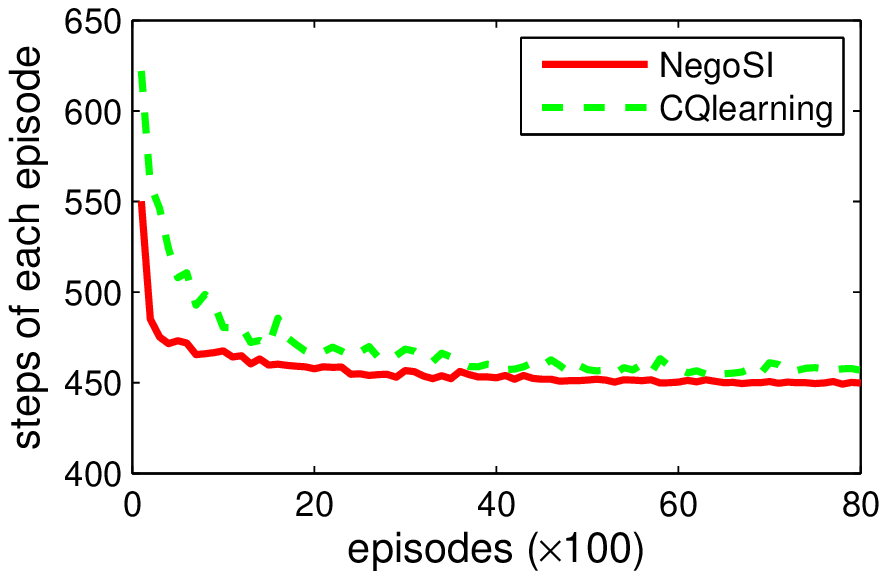}}
  \centerline{(a) SEE}
\end{minipage}
\hfill
\begin{minipage}{0.49\linewidth}\label{fig15b}
  \centerline{\includegraphics[width=2.6in]{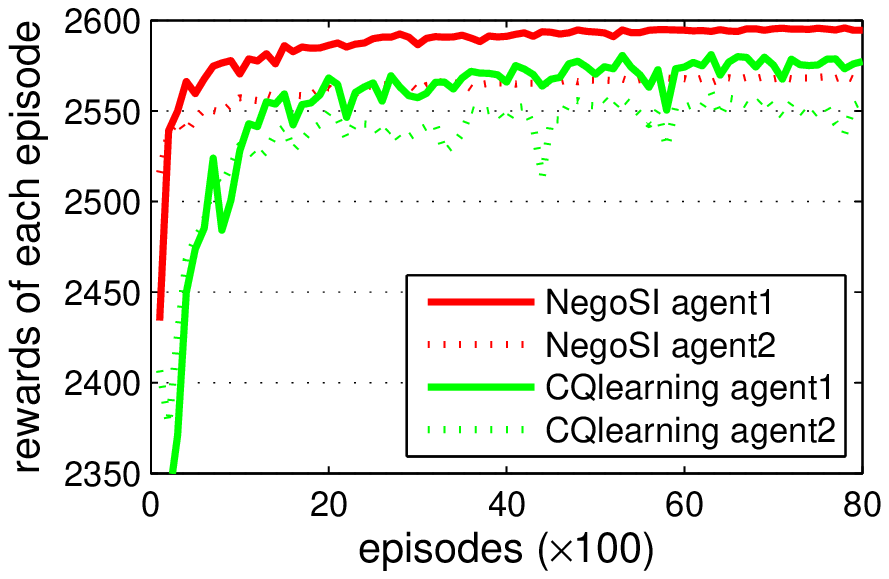}}
  \centerline{(b) REE}
\end{minipage}
\caption{SEE (steps of each episode) and REE (rewards of each episode) for NegoSI and CQ-learning in the 2-agent intelligent warehouse.}
\label{fig15}
\end{figure*}

\begin{figure*}[!bp]
\begin{minipage}{0.49\linewidth}\label{fig16a}
  \centerline{\includegraphics[width=2.6in]{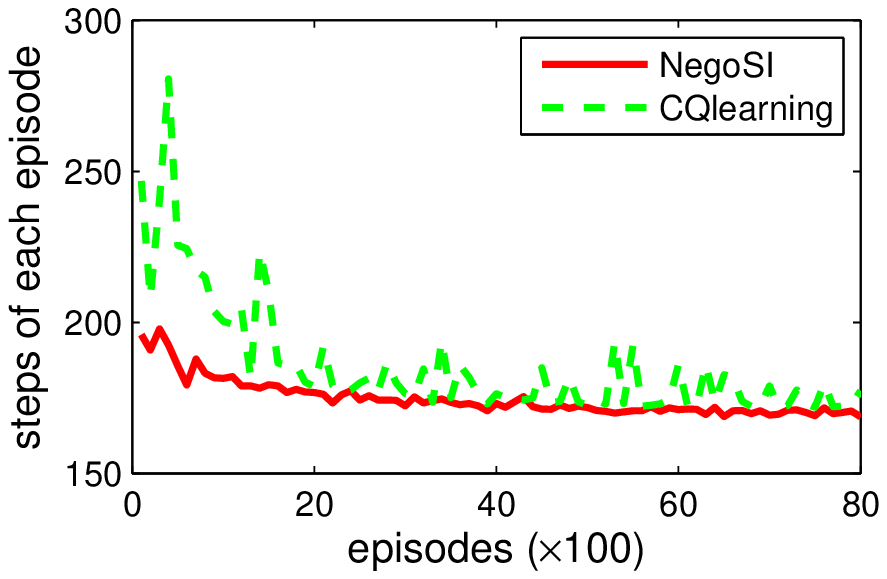}}
  \centerline{(a) SEE}
\end{minipage}
\hfill
\begin{minipage}{0.49\linewidth}\label{fig16b}
  \centerline{\includegraphics[width=2.6in]{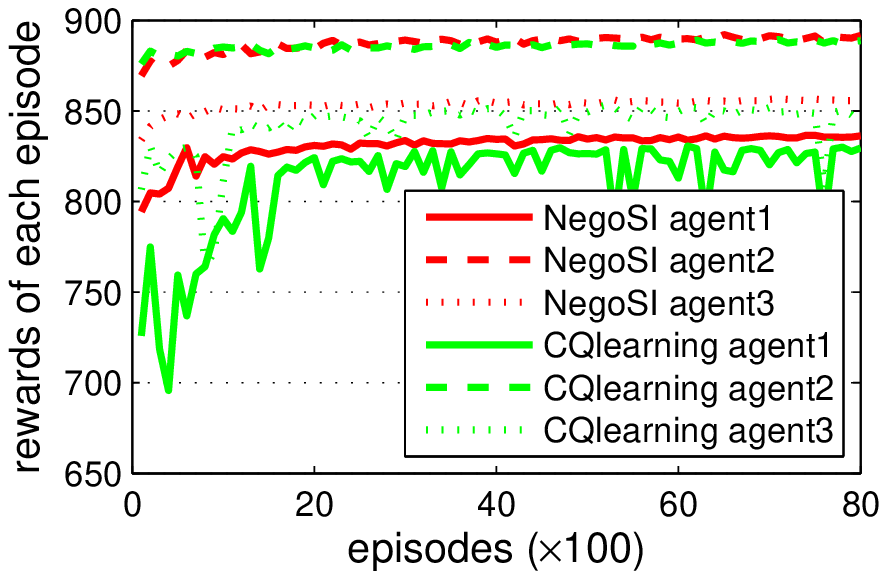}}
  \centerline{(b) REE}
\end{minipage}
\caption{SEE (steps of each episode) and REE (rewards of each episode) for NegoSI and CQ-learning in the 3-agent intelligent warehouse.}
\label{fig16}
\end{figure*}

We only compare the NegoSI algorithm with CQ-learning for the intelligent warehouse problem. In fact, ILVFT has no guarantee of the convergence characteristics and it is difficult to converge in the intelligent warehouse setting in practice. NegoVFT is also infeasible since its internal memory cost in MATLAB is estimated to be $208\times208\times208\times10\times10\times10\times4\times4\times4\times8B=4291GB$, while the costs of other algorithms are about 2 MB. Like NegoVFT, other MG-based MARL algorithms cannot solve the intelligent warehouse problem either. Experiments were performed in 2-agent and 3-agent settings, respectively, and the results are shown as in Fig. \ref{fig15} and Fig. \ref{fig16}.

In the 2-agent setting, the initial position and final goal of robot $1$ are ($1, 1$) and those of robot $2$ are ($1, 16$). Each robot needs to finish 30 randomly assigned tasks. The task set is the same for all algorithms. Robots with NegoSI achieve lower SEE (steps of each episode) than robots with CQ-learning throughout the whole learning process (as shown in Fig. \ref{fig15}). NegoSI finally converges to 449.9 steps and CQ-learning converges to 456.9 steps. In addition, robots with NegoSI have higher and more stable REE (rewards of each episode). Finally, the average runtime for completing all the tasks is $2227$s for NegoSI and is $3606$s for CQ-learning. Robots used 38\% less time to complete all the tasks with NegoSI than that with CQ-learning.

The performances of NegoSI are also better in the 3-agent setting than that of CQ-learning. The initial position and the final goal of different robots are ($1, 1$), ($1, 8$) and ($1, 16$). Each robot needs to finish $10$ randomly assigned tasks. The task set is the same for different algorithms. SEE (steps of each episode) for robots with NegoSI finally converges to $168.7$ steps and that for robots with CQ-learning converges to $177.3$ steps (as shown in Fig. \ref{fig16}). In addition, the learning curves of NegoSI are more stable. Robots with NegoSI have higher and more stable REE (rewards of each episode). Finally, the average runtime for completing all the tasks is $1352$s for NegoSI and $2814$s for CQ-learning. The robots use 52\% less time to complete all the tasks with NegoSI than that with CQ-learning.

\begin{remark}
The agents with CQ-learning algorithm learn faster than the agents
with NegoSI for the benchmark maps in Section \ref{Sec4.1}, but
slower than those with NegoSI for the intelligent warehouse
problem in Section \ref{Sec4.2}. The reason is that for agents
with CQ-learning, the number of ``coordination state" is several
times higher than that with NegoSI. This difference becomes
significant with the increase of the task number, the environment
scale and the number of agents. Thus, the time used to search for
one specific ``coordination state" in the ``coordination state"
pool increases faster in CQ-learning than in NegoSI, which results
in the increase of the whole learning time. According to all these
experimental results, the presented NegoSI algorithm maintains
better performances regarding such characteristics as coordination
ability, convergence, scalability and computational complexity,
especially for practical problems.
\end{remark}

\section{Conclusions}\label{Sec5}

In this paper a negotiation-based MARL algorithm with sparse
interactions (NegoSI) is proposed for the learning and
coordination problems in multi-agent systems. In this integrated
algorithm the knowledge transfer mechanism is also adopted to
improve agent's learning speed and coordination ability. In
contrast to traditional sparse-interaction based MARL algorithms,
NegoSI adopts the equilibrium concept and makes it possible for
agents to select non-strict EDSP or Meta equilibrium for their
joint actions, which makes it easy to find near optimal (or even
optimal) policy and to avoid collisions as well. The experimental
results demonstrate the effectiveness of the presented NegoSI
algorithm regarding such characteristics as fast convergence, low
computational complexity and high scalability in comparison to the
state-of-the-art MARL algorithms, especially for practical
problems. Our future work focuses on further comparison of NegoSI
with other existing MARL algorithms and more applications of MARL
algorithms to general and realistic problems. In addition,
multi-objective reinforcement learning (MORL) \cite{Vamplew and
Dazeley 2011}-\cite{Liu and Xu 2015} will also be considered to
further combine environment information and coordination knowledge
for local learning.

\section*{Acknowledgement}
The authors would like to thank Dr. Yujing Hu and Dr. Jiajia Dou for helpful discussion. The authors would also like to thank the Associate Editor and
the anonymous reviewers for their constructive comments that have greatly improved the original manuscript.


\begin{thebibliography}{99}

\bibitem{Busoniu and Babuska 2008}
L. Bu\c{s}oniu, R. Babu\v{s}ka and B. D. Schutter, ``A comprehensive survey of multi-agent reinforcement learning," \emph{IEEE Transactions on System, Man, and Cybernetics, Part C: Applications and Reviews}, vol. 38, no. 2, pp. 156-172, 2008.

\bibitem{Vrancx and Verbeeck 2008}
P. Vrancx, K. Verbeeck and A. Now\'e, ``Decentralized learning in markov games," \emph{IEEE Transactions on Systems, Man, and Cybernetics, Part B: Cybernetics}, vol. 38, no. 4, pp. 976-981, 2008.

\bibitem{Bianchi and Martins 2014}
R. Bianchi, M. F. Martins, C. H. Ribeiro and A. H. Costa, ``Heuristically-accelerated multiagent reinforcement learning," \emph{IEEE Transactions on Cybernetics}, vol. 44, no. 2, pp. 252-265, 2014.

\bibitem{Hwang and Tan 2005}
K. Hwang, S. Tan, M. Hsiao and C. Wu, ``Cooperative multiagent congestion control for high-speed networks," \emph{IEEE Transactions on Systems, Man, and Cybernetics, Part B: Cybernetics}, vol. 35, no. 2, pp. 255-268, 2005.

\bibitem{Yu and Zhang 2015}
C. Yu, M. Zhang, F. Ren and G. Tan, ``Multiagent learning of coordination in loosely coupled multiagent systems," \emph{IEEE Transaction on Cybernetics}, vol. 45, no. 12, pp. 2853-2867, 2015.

\bibitem{Littman 1994}
M. L. Littman, ``Markov games as a framework for multi-agent reinforcement learning," \emph{in Proceedings of the International Conference on Machine Learning (ICML-94)}, pp.157-163, New Brunswick, NJ, July 10-13, 1994.

\bibitem{Melo and Veloso 2011}
F. S. Melo and M. Veloso, ``Decentralized MDPs with sparse interactions," \emph{Artificial Intelligence}, vol. 175, no. 11, pp. 1757-1789, 2011.

\bibitem{Hu and Gao 2014-1}
Y. Hu, Y. Gao and B. An, ``Multiagent reinforcement learning with unshared value functions," \emph{IEEE Transaction on Cybernetics}, vol. 45, no. 4, pp. 647-662, 2015.

\bibitem{Hu and Gao 2014-2}
Y. Hu, Y. Gao and B. An, ``Accelerating multiagent reinforcement learning by equilibrium transfer," \emph{IEEE Transaction on Cybernetics}, vol.45, no. 7, pp. 1289 - 1302, 2015.

\bibitem{Hu and Wellman 2003}
J. Hu and M. P. Wellman, ``Nash Q-learning for general-sum stochastic games," \emph{The Journal of Machine Learning Research}, vol. 4, pp. 1039¨C1069, 2003.

\bibitem{Littman 2001}
M. L. Littman, ``Friend-or-foe Q-learning in general-sum games," \emph{in Proceedings of the International Conference on Machine Learning (ICML-01)}, pp. 322-328, Williams College, Williamstown, MA, USA, 2001.

\bibitem{Greenwald and Hall 2003}
A. Greenwald, K. Hall and R. Serrano, ``Correlated Q-learning," \emph{in Proceedings of the International Conference on Machine Learning (ICML-03)}, pp. 84-89, Washington, DC, USA, 2003.

\bibitem{Melo and Veloso 2009}
F. S. Melo and M. Veloso, ``Learning of coordination: Exploiting sparse interactions in multiagent systems," \emph{in Proceedings of the International Conference on Autonomous Agents and Multiagent Systems (AAMAS)}, pp. 773-780, 2009.

\bibitem{Hauwere and Vrancx 2010}
Y. D. Hauwere, P. Vrancx and A. Now\'e, ``Learning multi-agent state space representations," \emph{in Proceedings of the International Conference on Autonomous Agents and Multiagent Systems (AAMAS)}, vol. 1, no. 1, pp. 715-722, 2010.

\bibitem{Hauwere and Vrancx 2010-2}
Y. D. Hauwere, P. Vrancx and A. Now\'e, ``Generalized learning
automata for multi-agent reinforcement learning," \emph{AI
Communications}, vol. 23, no. 4, pp. 311-324, 2010.

\bibitem{Wang and Jiang 2014}
W. Wang and Y. Jiang, ``Community-aware task allocation for social networked multiagent systems," \emph{IEEE Transactions on Cybernetics}, vol. 44, no. 9, pp. 1529-1543, 2014.

\bibitem{Enright and Wurman 2011}
J. Enright and P. R. Wurman, ``Optimization and coordinated autonomy in mobile fulfillment systems," \emph{Automated Action Planning for Autonomous Mobile Robots}, pp. 33-38, 2011.

\bibitem{Stone and Veloso 2000}
P. Stone and M. Veloso, ``Multiagent systems: a survey from the machine learning perspective," \emph{Autonomous Robots}, vol. 8, no. 3, pp. 345-383, 2000.

\bibitem{Nash 1950}
J. F. Nash, ``Equilibrium points in n-person games," \emph{in Proceedings of the National Academy of Sciences of the United States of America}, vol. 36, no. 1, pp. 48-49, 1950.

\bibitem{An and Douglis 2008}
B. An, F. Douglis and F. Ye, ``Heuristics for negotiation schedules in multi-plan optimization," \emph{in Proceedings of the International Conference on Autonomous Agents and Multiagent Aystems (AAMAS)}, vol. 2, pp. 551-558, 2008.

\bibitem{An and Lesser 2011}
B. An, V. Lesser, D. Westbrook and M. Zink, ``Agent-mediated multi-step optimization for resource allocation in distributed sensor networks," \emph{in Proceedings of the International Conference on Autonomous Agents and Multiagent Systems (AAMAS)}, vol. 2, pp. 609-616, 2011.

\bibitem{Sutton and Barto 1998}
R. S. Sutton and A. G. Barto, ``Reinforcement learning: an introduction," \emph{MIT press}, 1998.

\bibitem{Watkins 1989}
C. Watkins, ``Learning from delayed rewards," PhD thesis, University of Cambridge, 1989.

\bibitem{Tsitsiklis 1994}
J. Tsitsiklis, ``Asynchronous stochastic approximation and Q-learning," \emph{Machine Learning}, vol. 16, no. 3, pp. 185-202, 1994.

\bibitem{Nowe and Vrancx 2012}
A. Now\'e, P. Vrancx, and Y. D. Hauwere, ``Game theory and multi-agent reinforcement learning," \emph{Reinforcement Learning}, Springer Berlin Heidelberg, pp. 441-470, 2012.

\bibitem{Burkov 2014}
A. Burkov and B. Chaib-Draa, ``Repeated games for multiagent systems: a survey," \emph{The Knowledge Engineering Review}, vol. 29, no. 1, pp. 1-30, 2014.

\bibitem{Burkov 2010}
A. Burkov and B. Chaib-Draa, ``An approximate subgame-perfect equilibrium computation technique for repeated games," \emph{Proceedings of the Twenty-Fourth AAAI Conference on Artificial Intelligence}, pp. 729-736, 2010.


\bibitem{Porter and Nudelman 2008}
R. Porter, E. Nudelman and Y. Shoham, ``Simple search methods for
finding a Nash equilibirum," \emph{Games and Economic Behavior,}
vol. 63, no. 2, pp. 642-662, 2008.

\bibitem{Hu and Gao 2015}
Y. Hu, Y. Gao and B. An, ``Learning in Multi-agent Systems with
Sparse Interactions by Knowledge Transfer and Game Abstraction,"
\emph{in Proceedings of the International Conference on Autonomous
Agents and Multiagent Systems (AAMAS)}, pp. 753-761, Istanbul,
Turkey, 4-8 May, 2015.

\bibitem{Guestrin and Lagoudakis 2002}
C. Guestrin, M. Lagoudakis and R. Parr, ``Coordinated reinforcement learning," \emph{in Proceedings of the International Conference on Machine Learning (ICML-02)}, vol. 2, pp. 227-234. 2002.

\bibitem{Guestrin and Venkataraman 2002}
C. Guestrin, S. Venkataraman and D. Koller, ``Context-specific multiagent coordination and planning with factored MDPs," \emph{in Proceedings of the National Conference on Artificial Intelligence}, pp. 253-259. 2002.

\bibitem{Kok and Vlassis 2004}
J. R. Kok and N. A. Vlassis, ``Sparse cooperative Q-learning," \emph{in Proceedings of the International Conference on Machine Learning (ICML-04)}, pp. 61-68, Banff, Alberta, Canada, 2004.

\bibitem{Hauwere and Vrancx 2012}
Y. D. Hauwere, P. Vrancx, and A. Now\'e, ``Solving sparse delayed coordination problems in multi-agent reinforcement learning," \emph{Adaptive and Learning Agents}, Springer Berlin Heidelberg, pp. 114-133, 2012.

\bibitem{Vrancx and Hauwere 2011}
P. Vrancx, Y. D. Hauwere and A. Now\'e, ``Transfer learning for multi-agent coordination," \emph{in Proceedings of the International Conference on Agents and Artificial Intelligence (ICAART)}, pp. 263-272, 2011.

\bibitem{Claus and Boutilier 1998}
C. Claus and C. Boutilier, ``The dynamics of reinforcement learning in cooperative multiagent systems," \emph{in Proceedings of the fifteenth national/tenth conference on Artificial intelligence/Innovative applications of artificial intelligence}, pp. 746-752, July 26¨C30, 1998.

\bibitem{Liu and Yang 2015}
D. Liu, X. Yang, D. Wang and Q. Wei,
``Reinforcement-learning-based robust controller design for
continuous-time uncertain nonlinear systems subject to input
constraints," \emph{IEEE Transactions on Cybernetics}, vol. 45,
no. 7, pp. 1372-1385, 2015.

\bibitem{Haeri and Trajkovic 2015}
S. Haeri and L. Trajkovic, ``Intelligent Deflection Routing in Buffer-Less Networks," \emph{IEEE Transactions on Cybernetics}, vol. 45, no. 2, pp. 316-327, 2015.

\bibitem{Ernst and Glavic 2009}
D. Ernst, M. Glavic, F. Capitanescu and L. Wehenkel, ``Reinforcement learning versus model predictive control: a comparison on a power system problem," \emph{IEEE Transactions on Systems, Man, and Cybernetics, Part B: Cybernetics}, vol. 39, no. 2, pp. 517-529, 2009.

\bibitem{Valasek and Doebbler 2008}
J. Valasek, J. Doebbler, M. D. Tandale and A. J. Meade, ``Improved adaptive-reinforcement learning control for morphing unmanned air vehicles," \emph{IEEE Transactions on Systems, Man, and Cybernetics, Part B: Cybernetics}, vol. 38, no. 4, pp. 1014-1020, 2008.

\bibitem{Sabattini and Secchi 2015}
L. Sabattini, C. Secchi and N. Chopra, ``Decentralized estimation
and control for preserving the strong connectivity of directed
graphs," \emph{IEEE Transactions on Cybernetics}, vol. 45, no. 10,
pp. 2273-2286, 2015.

\bibitem{Su and Lin 2016}
S. Su, Z. Lin and A. Garcia, ``Distributed synchronization control
of multiagent systems with unknown nonlinearities," \emph{IEEE
Transactions on Cybernetics}, vol. 46, no. 1, pp. 325-338, 2016.

\bibitem{Zhang and Feng 2015}
H. Zhang, T. Feng, G. Yang and H. Liang, ``Distributed cooperative
optimal control for multiagent systems on directed graphs: an
inverse optimal approach," \emph{IEEE Transactions on
Cybernetics}, vol. 45, no. 7, pp. 1315-1326, 2015.

\bibitem{Zhou and Shi 2014}
L. Zhou, Y. Shi, J. Wang and P. Yang, ``A balanced heuristic mechanism for multirobot task allocation of intelligent warehouses," \emph{Mathematical Problems in Engineering}, vol. 2014, article ID 380480, 2014.

\bibitem{Viet and Kyaw 2011}
H. H. Viet, P. H. Kyaw and T. Chung. ``Simulation-based evaluations of reinforcement learning algorithms for autonomous mobile robot path planning," \emph{IT Convergence and Services}, Springer Netherlands, pp. 467-476, 2011.

\bibitem{Wurman and D'Andrea 2008}
P. R. Wurman, R. D'Andrea and M. Mountz, ``Coordinating hundreds of cooperative, autonomous vehicles in warehouses," \emph{AI Magazine}, vol. 29, no. 1, pp. 9-20, 2008.

\bibitem{Vamplew and Dazeley 2011}
P. Vamplew, R. Dazeley, A. Berry, R. Issabekov and E. Dekker, ``Empirical evaluation methods for multiobjective reinforcement learning algorithms," \emph{Machine Learning}, vol. 84, no. 1-2, pp. 51-80, 2011.

\bibitem{Liu and Chung 1999}
C. Lin and I. Chung, ``A reinforcement neuro-fuzzy combiner for multiobjective control," \emph{IEEE Transactions on Systems, Man, and Cybernetics, Part B: Cybernetics}, vol. 29, no. 6, pp. 726-744, 1999.

\bibitem{Liu and Xu 2015}
C. Liu, X. Xu and D. Hu, ``Multiobjective reinforcement learning: A comprehensive overview," \emph{IEEE Transactions on Systems, Man, and Cybernetics: Systems}, vol. 45, no. 3, pp. 385-398, 2015.

\end{thebibliography}
\end{document}